\documentclass[twocolumn,prb,reprint,amsmath,amssymb,aps,showpacs]{revtex4-1}

\usepackage{graphicx}
\usepackage{dcolumn}
\usepackage{bm}
\usepackage{graphics}
\usepackage{epsfig}
\usepackage{epstopdf}
\begin{document}

\title{Spontaneous symmetry breaking and coherence in two-dimensional electron-hole and exciton systems}

\author{S.A. Moskalenko,$^{1}$ M.A. Liberman,$^{2}$ E.V. Dumanov,$^{1}$
and E.S. Moskalenko$^{3}$}
\affiliation{$^{1}$Institute of Applied Physics of the Academy of Sciences of Moldova, Academic Str. 5, Chisinau, MD2028, Republic of Moldova\\
$^{2}$Department of Physics, Uppsala University, Box 530, SE-751 21, Uppsala, Sweden\\
$^{3}$A F Ioffe Physical-Technical Institute, Russian Academy of Sciences, 26 Politekhnicheskaya, 194021 St Petersburg, Russia}

\date{\today}

\begin{abstract}
The spontaneous breaking of the continuous symmetries of the two-dimensional (2D) electron-hole systems in a strong perpendicular magnetic field leads to the formation of new ground states and determines the energy spectra of the collective elementary excitations appearing over these ground states. In this review the main attention is given to the electron-hole systems forming coplanar magnetoexcitons in the Bose-Einstein condensation (BEC) ground state with the wave vector $\vec{k}=0$, taking into account the excited Landau levels, when the exciton-type elementary excitations coexist with the plasmon-type oscillations. At the same time properties of the two-dimensional electron gas (2DEG) spatially separated as in the case of double quantum wells (DQWs) from the 2D hole gas under conditions of the fractional quantum Hall effect (FQHE) are of great interest because they can influence the quantum states of the coplanar magnetoexcitons when the distance between the DQW layers diminishes. We also consider in this review the bilayer electron systems under conditions of the FQHE with the one half filling factor for each layer and with the total filling factor for two layers equal to unity because the coherence between the electron states in two layers is equivalent to the formation of the quantum Hall excitons (QHExs) in a coherent macroscopic state. This makes it possible to compare the energy spectrum of the collective elementary excitations of the Bose-Einstein condensed QHExs and coplanar magnetoexcitons. The breaking of the global gauge symmetry as well as of the continuous rotational symmetries leads to the formation of the gapless Nambu-Goldstone (NG) modes while the breaking of the local gauge symmetry gives rise to the Higgs phenomenon characterized by the gapped branches of the energy spectrum. These phenomena are equivalent to the emergence of massless and of massive particles, correspondingly, in the relativistic physics. The application of the Nielsen-Chadha theorem establishing the number of the NG modes depending of the number of the broken symmetry operators and the elucidation when the quasi-NG modes appear are demonstrated using as an example related with the BEC of spinor atoms in an optical trap. They have the final aim to better understand the results obtained in the case of the coplanar Bose-Einstein condensed magnetoexcitons. The Higgs phenomenon results in the emergence of the composite particles under the conditions of the FQHE. Their description in terms of the Ginzburg-Landau theory is remembered. The formation of the high density 2D magnetoexcitons and magnetoexciton-polaritons with point quantum vortices attached is suggested.
 The conditions in which the spontaneous coherence could appear in a system of indirect excitons in a double quantum well structures are discussed. The experimental attempts to achieve these conditions, the main results and the accumulated knowledge are reviewed.
\end{abstract}

\maketitle
\tableofcontents


\section{Introduction}

The collective elementary excitations of the two-dimensional (2D) electron-hole (e-h) systems in a strong perpendicular magnetic field are discussed in the frame of the Bogoliubov theory of quasiaverages [1] taking into account the phenomena related with the spontaneous breaking of the continuous symmetries. The main results in this field have been obtained thanks to the fundamental papers by Goldstone [2], Nambu [3], Higgs [4] and Weinberg [5]. These investigations were influenced by the success of the theory of superconductivity developed originally by Bardin, Cooper and Schriffer [6], refined later by Bogoliubov [1] as well as by the microscopic theory of superfluidity proposed by Bogoliubov [1]. The specific implementation of these concepts and theorems in the case of 2D magnetoexcitons with direct implication of the plasmon-type excitations side-by-side with the exciton-type branches of the energy spectrum is the main topic of the present review. The coplanar electrons and holes in a strong perpendicular magnetic field at low temperatures form the magnetoexcitons, when the Coulomb interaction between electrons and holes lying on the lowest Landau levels (LLLs) plays the main role. However, when the electrons and holes are spatially separated on the different layers of the double quantum well (DQW) the Coulomb e-h interaction diminishes, and the two-dimensional electron gas (2DEG) on one layer and the two-dimensional hole gas (2DHG) on another layer are formed. Their properties under the conditions of the fractional quantum Hall effect (FQHE) can influence the properties of the 2D magnetoexcitons. To the best of our knowledge these aspects of the magnetoexciton physics were not discussed in literature.
 A short review is given on the Bose-Einstein Condensation (BEC) of the quantum Hall excitons (QHExs) arising in the bilayer electron systems under the conditions of the FQHE at one half filling factor for each layer and the total filling factor equal to unity for both layers. This enables us to compare the phenomenon of the BEC of coplanar magnetoexcitons and of QHExs. Such comparison provides better understanding of the underlying physics and allows to verify accuracy of the made approximations. Because the point vortices play an important role in the understanding of the FQHE the corresponding additional information should be included. The possibility to consider the BEC at $T=0$ as an estimate for the finite temperatures below the Berezinskii-Kosterlitz-Thouless phase transition is suggested.
 The article is organized as follow. In Section 2 the Bogoliubov theory of the quasiaverages is overviewed. Section 3 is devoted to the Goldstone theorem. The Nambu-Goldstone modes arising under the condition of BEC of the sodium atoms are enumerated in Section 4. The breaking of the local gauge symmetry and the Higgs phenomenon are discussed in Section 5. Section 6 is devoted to the quasi-Nambu-Goldstone modes. In Section 7 the Ginzburg-Landau theory for the FQHE is formulated. The 2D point quantum vortices are described in Section 8. The existence of the statistical gauge vector potential generated by the vortices is considered in Section 9. The BEC of QHExs and the energy spectrum of elementary excitations under these conditions are discussed in Section 10. Section 11 contains the main results concerning the energy spectrum of the exciton and plasmon branches of the collective elementary excitations of the Bose-Einstein condensed coplanar magnetoexcitons.
\section{Bogoliubov's Theory of Quasiaverages}
N.N. Bogoliubov [1] has demonstrated his concept of quasiaverages using the ideal Bose-gas model with the Hamiltonian
\begin{equation}
H=\sum\limits_{k}\left( \dfrac{\hbar ^{2}k^{2}}{2m}-\mu \right) a_{k}^{\dag
}a_{k},\hspace{5mm}
\end{equation}
here $a_{k}^{\dag } ,\text{\ } a_{k} $ are the Bose operators of creation
and annihilation of particles, and $\mu $ is their chemical potential.
\newline
The occupation numbers of the particles are
\begin{equation}
N_{0}=\dfrac{1}{e^{-\beta \mu }-1};\text{\ }N_{k}=\dfrac{1}{e^{\beta \left(
\dfrac{\hbar ^{2}k^{2}}{2m}-\mu \right) }-1}
\end{equation}
where $\mu \leq 0$ and $\beta =1/kT$. \newline
In the normal state, the density of particles in the thermodynamic limit at $%
\mu =0$ becomes $n=2.612\left( mk_{B} T\right) ^{3/2} /\left( 2\pi \hbar
^{2} \right) ^{3/2} $. At this point, the Bose-Einstein condensation occurs
and a finite value of the density of condensed particles appears in the
thermodynamic limit
\begin{equation}
n_{0}=\lim\limits_{V\rightarrow \infty }\dfrac{N_{0}}{V};\text{\ }\mu
=-k_{B}T\ln \left( 1+\dfrac{1}{N_{0}}\right)
\end{equation}
The operators $a_{0}^{\dag }$ and $a_{0}$ asymptotically become $c-$numbers,
when their commutator \newline
\begin{equation}
\left[ \dfrac{a_{0}}{\sqrt{V}},\dfrac{a_{0}^{\dag }}{\sqrt{V}}\right] =%
\dfrac{1}{V}
\end{equation}
asymptotically tends to zero and their product is equal to $n_{0}$. Then one
can write
\begin{equation}
\dfrac{a_{0}^{\dag }}{\sqrt{V}}\sim \sqrt{n_{0}}e^{i\alpha };\text{\ \ }%
\dfrac{a_{0}^{{}}}{\sqrt{V}}\sim \sqrt{n_{0}}e^{-i\alpha }
\end{equation}
On the other hand, the regular averages of the operators $a_{0}^{\dag }$ and $a_{0}$
in the Hamiltonian (1) are exactly equal to zero. It is the consequence of
the commutativity of the operator $H$ and the operator of the total particle
number $N$ as follows
\begin{equation}
\hat{N}=\sum\limits_{k}a_{k}^{\dag }a_{k};\text{\ \ }[H,\hat{N}]=0.
\end{equation}
As a result, the operators $H$ is invariant with respect to the unitary
transformation
\begin{equation}
U=e_{{}}^{i\hat{N}\phi }
\end{equation}
with an arbitrary angle $\phi $. This invariance is called
gradient invariance of the first kind or gauge invariance. When $\phi $ does
not depend on the coordinate $x$, we have the global gauge invariance and in
the case $\phi (x)$ it is called local gauge invariance [2-8] or gauge
invariance of the second kind.
The invariance (7) implies $H=U^{\dag } HU;\text{\ \ } U^{\dag} a_{0} U=e^{i\phi } a_{0} $, which leads to the following average value
\begin{gather*}
\left\langle a_{0} \right\rangle \cong Tr\left( a_{0} e^{-\beta H} \right)
=Tr\left( a_{0} Ue^{-\beta H} U^{\dag } \right) =\\
=Tr\left( U^{\dag } a_{0}Ue^{-\beta H} \right) =e^{i\phi } \left\langle a_{0} \right\rangle ; \\
\left( 1-e^{i\phi } \right) \left\langle a_{0} \right\rangle =0
\end{gather*}
Because $\phi $ is an arbitrary angle, there are the selection rules:
\begin{equation}
\left\langle a_{0}\right\rangle =0;\text{\ \ }\left\langle a_{0}^{\dag
}\right\rangle =0
\end{equation}
The regular average (8) can also be obtained from the
asymptotical expressions (5) if they are integrated over the angle $\alpha$.
This apparent contradiction can be resolved if Hamiltonian (1) is
supplemented by additional term
\begin{equation}
-\nu \left( a_{0}^{\dag }e^{i\varphi }+a_{0}e^{-i\varphi }\right) \sqrt{V},%
\text{\ }\nu >0,
\end{equation}
where $\varphi $ is the fixed angle and $\nu $ is infinitesimal value.

New Hamiltonian has the form
\begin{equation}
H_{\nu ,\phi }=\sum\limits_{k}\left( \dfrac{\hbar ^{2}k^{2}}{2m}-\mu \right)
a_{k}^{\dag }a_{k}-\nu \left( a_{0}^{\dag }e^{i\varphi }+a_{0}e^{-i\varphi
}\right) \sqrt{V}.
\end{equation}
It does not conserve the condensate number. Now the regular average values of the
operators $a_{0}^{\dag }$ and $a_{0}$ over the Hamiltonian $H_{\nu ,\phi }$
differ from zero, i.e., $\left\langle a_{0}\right\rangle _{H\varphi }\neq 0$
and $\left\langle a_{0}^{\dag }\right\rangle _{H\varphi }\neq 0$. The
definition of the quasiaverages designated by $<a_{0}>$ is the limit of the
regular average $\left\langle a_{0}\right\rangle _{H\varphi }$ when $\nu $
tends to zero
\begin{equation}
<a_{0}>=\lim\limits_{\nu \rightarrow 0}\left\langle a_{0}\right\rangle
_{H_{\nu ,\varphi }}.
\end{equation}
It is important to emphasize that the limit $\nu \rightarrow 0$
must be effectuated after the thermodynamic limit $V\rightarrow \infty $, $%
N_{0}\rightarrow \infty $. In the thermodynamic limit, $\mu $ is also
infinitesimal, and it is possible to choose the ratio of two infinitesimal
values $\mu $ and $\nu $ obtaining a finite value
\begin{equation}
-\dfrac{\nu }{\mu }=\sqrt{n_{0}}.
\end{equation}
To calculate the regular average $\left\langle
a_{0}\right\rangle _{H_{\nu ,\varphi }}$ one needs to represent the
Hamiltonian (10) $H_{\nu ,\phi }$ in a diagonal form with the aid of the
canonical transformation over the amplitudes
\begin{eqnarray}
a_{0} &=&-\dfrac{\nu }{\mu }e^{i\varphi }\sqrt{V}+\alpha _{0}; \\
a_{k} &=&\alpha _{k};\text{ }k\neq 0.  \notag
\end{eqnarray}
In terms of
the new variables the Hamiltonian $H_{\nu ,\phi }$ takes the form
\begin{equation}
H_{\nu ,\phi }=-\mu \alpha _{0}^{\dag }\alpha _{0}+\sum\limits_{k}\left(
\dfrac{\hbar ^{2}k^{2}}{2m}-\mu \right) \alpha _{\vec{k}}^{\dag }\alpha _{%
\vec{k}}+\dfrac{\nu ^{2}V}{\mu }.
\end{equation}
In the diagonal representation (14), the regular average value $\left\langle \alpha
_{0}\right\rangle _{H_{\nu ,\varphi }}$ exactly equals to zero, while the
value $\left\langle a_{0}\right\rangle _{H_{\nu ,\varphi }}$ is equal to the
first term on the right-hand side of formulas (13).

As a result, the quasiaverage $<a_{0} >$ is
\begin{equation}
<a_{0}>=\lim\limits_{\nu \rightarrow 0}\left\langle a_{0}\right\rangle
_{H_{\nu ,\varphi }}=\sqrt{N_{0}}e^{i\varphi }
\end{equation}
It depends on the fixed angle $\varphi $ and does not depend on
$\nu $. The spontaneous global gauge symmetry breaking is implied when the
phase $\varphi $ of the condensate amplitude in Hamiltonian (10) is fixed.

When the interaction between the particles is taken into
account, these differences appear for other amplitudes as well. They give
rise to the renormalization of the energy spectrum of the collective
elementary excitations. In such a way, the canonical transformation
\begin{equation}
a_{k}=\sqrt{N_{0}}\delta _{k,0}e^{i\varphi }+\alpha _{k}
\end{equation}
introduced for the first time by Bogoliubov [1] in his theory
of superfluidity, has a quantum-statistical foundation within the framework
of the quasiaverage concept. At $T=0$ the quasiaverage $<a_{0}>$ coincides
with the average over the quantum-mechanical ground state, which is the
coherent macroscopic state [9].

The phenomena related to the spontaneous breaking of the
continuous symmetry play an important role in statistical physics.\newline
Some elements of this concept, such as the coherent macroscopic state with a
given fixed phase and the displacement canonical transformation of the field
operator describing the Bose-Einstein condensate, were introduced by
Bogoliubov in the microscopical theory of superfluidity [1] and were
generalized in his theory of quasiaverages [1] noted above. \newline
The brief review of the gauge symmetries, their spontaneous breaking,
Goldstone and Higgs effects will be presented below following the Ryder's
monograph [7] and Berestetskii's lectures [8].
\section{Goldstone's Theorem}
Goldstone has considered a simple model of the complex scalar Bose field to
demonstrate his main idea. In the classical description the Lagrangian is
\begin{equation}
L=\left( \dfrac{\partial \phi ^{\ast }}{\partial x_{\mu }}\right) \left(
\dfrac{\partial \phi }{\partial x^{\mu }}\right) -m^{2}\phi ^{\ast }\phi
-\lambda \left( \phi ^{\ast }\phi \right) ^{2}
\end{equation}
The potential energy $V(\phi )$ has the form
\begin{equation}
V(\phi )=m^{2}\phi ^{\ast }\phi +\lambda \left( \phi ^{\ast }\phi \right)
^{2};\text{\ \ }\lambda >0,
\end{equation}
where $m^{2}$ is considered as a parameter only, rather than a mass term, $\lambda $ is
the parameter of self-interaction, whereas the denotations $x_{\mu }$ and $%
x^{\mu }$ mean
\begin{equation}
x^{\mu }=(ct,\vec{x});\text{\ \ }x_{\mu }=(ct,-\vec{x});
\end{equation}
The Lagrangian is invariant under the global gauge transformation
\begin{equation}
\phi =e^{i\Lambda }\phi ^{\prime };\text{\ \ }L(\phi )=L(\phi ^{\prime });%
\text{\ \ }\Lambda \text{-const}.
\end{equation}
It has the global gauge symmetry. The ground state is obtained by minimizing the
potential as follows\newline
\begin{equation}
\dfrac{\partial V(\phi )}{\partial \phi }=m^{2}\phi ^{\ast }+2\lambda \phi
^{\ast }\left\vert \phi \right\vert ^{2}.
\end{equation}
Of interest is the case $m^{2}<0$, when the minima are situated along the
ring
\begin{equation}
\left\vert \phi \right\vert ^{2}=-\dfrac{m^{2}}{2\lambda }=a^{2};\text{\ \ }%
\left\vert \phi \right\vert =a;\text{\ \ }a>0.
\end{equation}
The function $V(\phi )$ is shown in Fig. 1 being plotted against two real
components of the fields $\phi _{1}\text{\ and\ }\phi _{2}$.
\begin{figure}
\resizebox{0.48\textwidth}{!}{%
  \includegraphics{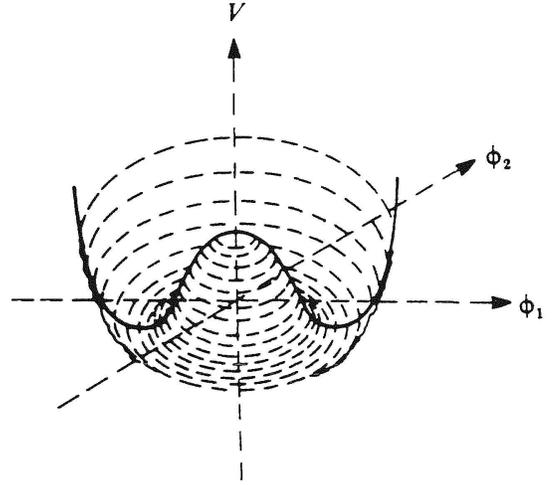}
}
\caption{ The potential $V(\phi)$ with the minima at $|\phi|=a$ and a local maximum at $\phi=0$}
\label{fig:1}       
\end{figure}
There is a set of degenerate vacuums related to each other by rotation. The
complex scalar field can be expressed in terms of two scalar real fields,
such as $\rho (x)$ and $\theta (x)$, in polar coordinates representation or
in the Cartesian decomposition as follows
\begin{equation}
\phi (x)=\rho (x)e^{i\theta (x)}=\left( \phi _{1}(x)+i\phi _{2}(x)\right)
\dfrac{1}{\sqrt{2}}.
\end{equation}
The Bogoliubov-type canonical transformation breaking the global gauge symmetry
is
\begin{equation}
\phi (x)=a+\dfrac{\phi _{1}^{\prime }(x)+i\phi _{2}^{\prime }(x)}{\sqrt{2}}%
=(\rho ^{\prime }(x)+a)e^{i\theta ^{\prime }(x)}.
\end{equation}
The new
particular vacuum state has the average $\left\langle \phi \right\rangle
_{0}=a$ with the particular vanishing vacuum expectation values $%
\left\langle \phi _{1}^{\prime }\right\rangle _{0}=\left\langle \phi
_{2}^{\prime }\right\rangle _{0}=\left\langle \rho ^{\prime }\right\rangle
_{0}=\left\langle \theta ^{\prime }\right\rangle _{0}=0$. It means the
selection of one vacuum state with infinitesimal phase $\theta ^{\prime
}\rightarrow 0$. As was pointed in [7], the physical fields are the
excitations above the vacuum. They can be realized by performing
perturbations about $\left\vert \phi \right\vert =a$. Expanding the
Lagrangian (17) in series of the infinitesimal perturbations $\theta
^{\prime }$, $\rho ^{\prime }$, $\phi _{1}^{\prime }$, $\phi _{2}^{\prime }$
and neglecting by the constant terms, we obtain
\begin{eqnarray}
L &=&\dfrac{1}{2}\left( \partial _{\mu }\phi _{1}^{\prime }\right) \left(
\partial ^{\mu }\phi _{1}^{\prime }\right) +\dfrac{1}{2}\left( \partial
_{\mu }\phi _{2}^{\prime }\right) \left( \partial ^{\mu }\phi _{2}^{\prime
}\right) -2\lambda a^{2}\phi _{1}^{\prime }{}^{2}-  \notag \\
&&-\sqrt{2}\lambda \phi _{1}^{\prime }\left( \phi ^{\prime }{}_{1}^{2}+\phi
_{2}^{\prime }{}^{2}\right) -\dfrac{\lambda }{4}\left( \phi _{1}^{\prime
}{}^{2}+\phi _{2}^{\prime }{}^{2}\right) ^{2}
\end{eqnarray}
or in polar description
\begin{gather}
L=\left( \partial _{\mu }\rho ^{\prime }\right) \left( \partial ^{\mu }\rho
^{\prime }\right) +\left( \rho ^{\prime }+a\right) ^{2}\left( \partial _{\mu
}\theta ^{\prime }\right) \left( \partial ^{\mu }\theta ^{\prime }\right) -
\notag \\
-\left[ \lambda \rho ^{\prime }{}^{4}+4a\lambda \rho ^{\prime
}{}^{3}+4\lambda a^{2}\rho ^{\prime }{}^{2}-\lambda a^{4}\right]
\end{gather}
Neglecting by the cubic and quartic terms, we will see that there are the
quadratic terms only of the type $4\lambda a^{2} \rho^{\prime}{}^{2}$ and $2\lambda a^{2} \phi^{\prime}_{1}{}^{2}$, but there is no quadratic terms
proportional to $\theta^{\prime}{}^{2} $ and $\phi^{\prime}_{2}{}^{2} $. For
real physical problems, for example, for the field theory, the field
components $\phi ^{\prime} _{1} $ and $\rho ^{\prime} $ represent massive
particles and dispersion laws with energy gap, whereas the field components $%
\phi ^{\prime} _{2} $ and $\theta ^{\prime} $ represent the massless
particles and gapless energy spectrum. \newline
The main Goldstone results can be formulated as follows
\begin{gather}
m_{\rho ^{\prime }}^{2}=4\lambda a^{2},\text{\ \ }m_{\phi _{1}^{\prime
}}^{2}=2\lambda a^{2};  \notag \\
m_{\theta ^{\prime }}^{2}=0;\text{\ \ }m_{\phi _{2}^{\prime }}^{2}=0;
\end{gather}
The spontaneous breaking of the global gauge symmetry takes place due to the
influence of the quantum fluctuations. They transform the initial field $%
\phi $ with two massive real components $\phi _{1} $ and $\phi _{2} $, and a
degenerate ground state with the minima forming a ring into another field
with one massive and other massless components, the ground state of which
has a well defined phase without initial symmetry.

The elementary excitations above the new ground state changing
the value $\left\langle \rho \right\rangle =a$ are massive. It costs energy
to displace $\rho ^{\prime} $ against the restoring forces of the potential $%
V(\rho )$. But there are no restoring forces corresponding to displacements
along the circular valley $\left| \phi \right| =a$ formed by initial
degenerate vacuums. \newline
Hence, for angular excitations $\theta ^{\prime} $of
wavelength, $\lambda $ we have $\omega \sim \lambda ^{-1} \rightarrow 0$ as $%
\lambda \rightarrow \infty $. The dispersion law is $\omega \sim ck$ and the
particles are massless [7]. The $\theta ^{\prime} $ particles are known as
the Goldstone bosons. This phenomenon is general and takes place in any
order of perturbation theory. The spontaneous breaking of a continuous
symmetry not only of the type as a global gauge symmetry but also of the
type of rotational symmetry entails the existence of massless particles
referred to as Goldstone particles or Nambu-Golstone gapless modes. This
statement is known as Goldstone theorem. It establishes that there exists a
gapless excitation mode when a continuous symmetry is spontaneously broken.
The angular excitations $\theta ^{\prime} $ are analogous to the spin waves.
The latter represent a slow spatial variation of the direction of
magnetization without changing of its absolute value. Since the forces in a
ferromagnetic are of short range, it requires a very little energy to excite
this ground state. So, the frequency of the spin waves has the dispersion
law $\omega =ck$. As was mentioned by Ryder [7], this argument breaks down
if there are long-range forces like, for example, the $1/r$ Coulomb force.
In this case, we deal with the maxwellian gauge field with local depending
on $x$ gauge symmetry instead of global gauge symmetry considered above.

After the specific application of the above statement will be
demonstrated following Refs. [10-16], where the spinor Bose-Einstein
condensates were discussed, we will consider the case of Goldstone field $%
\phi $ and of the maxwellian field with local gauge symmetry.
\section{Bogoliubov's Excitations and the Nambu-Goldstone Modes}
The above formulated theorems can be illustrated using the
specific example of the Bose-Einstein condensed sodium atoms $^{23}$Na in an
optical-dipole trap following the investigations of Murata, Saito and Ueda
[10] on the one side and of Uchino, Kobayashi and Ueda [11] on the other
side. There are numerous publications on this subject among which should be
mentioned [12-17]. The sodium atom $^{23}$Na has spin $f=1$ of the hyperfine
interaction and obey the Bose statistics. Resultant spin of the interacting
bosons with $f=1$ is $F$ which takes the values $F=0,1,2$. The contact
hard-core interaction constant $g_{F} =4\pi \hbar ^{2} a_{F} /M$ is
characterized by s-wave scattering length $a_{F} $, which is not zero for $%
F=0$ when two atomic spins form a singlet, and for $F=2$, when they form a
quintuplet. The constant $g_{0} $ and $g_{2} $ enter the combinations $c_{0}
=(g_{0} +2g_{2} )/3$ and $c_{1} =(g_{2} -g_{0} )/3$ which determine the
Hamiltonian. The description of the atomic Bose gas in an optical-dipole
trap is possible in the plane-wave representation due to the homogeneity and
the translational symmetry of the system. It means that the components of
the Bose field operator $\psi _{m} (\vec{r} )$ can be represented in the
form:
\begin{equation}
\psi _{m}(\vec{r})=\dfrac{1}{\sqrt{V}}\sum\limits_{k}a_{km}e^{i\vec{k}\vec{r}%
}
\end{equation}
where $V$ is the volume of the system, and $a_{km}$ is the annihilation
operator with the wave vector $\vec{k}$ and the magnetic quantum number $m$,
which in the case $f=1$ takes three values 1, 0, -1. The spinor
Bose-Einstein condensates were realized experimentally by the MIT group [12]
for different spin combinations using the sodium atoms $^{23}$Na in a
hyperfine spin states $|f=1,m_{f}=-1\rangle$ in a magnetic trap
and then transforming them to the optical-dipole trap formed by the single
infrared laser. The Bose-Einstein condensates were found to be long-lived.
Some arguments concerning the metastable long-lived states were formulated.
The states may appear if the energy barriers exist, which prevent the system
from direct evolving toward its ground states. If the thermal energy
needed to overcome these barriers is not available, the metastable state may
be long-lived and these events are commonly encountered. Even the
Bose-Einstein condensates in the dilute atomic gases can also be formed due
to the metastability. Moreover, in the gases with attractive interactions
the Bose-Einstein condensates may be metastable against the collapse just
due to the energy barriers [12]. Bellow we will discuss the Bogoliubov-type
collective elementary excitations arising over the metastable long-lived
ground states of the spinor-type Bose-Einstein condensates (BEC-tes)
following Ref.[10, 11], so as to demonstrate the formation of the
Nambu-Goldstone modes.

The Hamiltonian considered in [10] is given by formulas (3) and
(4), and has the form
\begin{eqnarray}
H &=&\sum\limits_{\vec{k},m}\left( \varepsilon _{\vec{k}}-pm+qm^{2}\right)
a_{\vec{k}m}^{\dag }a_{\vec{k}m}+  \notag \\
+\dfrac{c_{0}}{2V}\sum\limits_{\vec{k}} &:&\hat{\rho}_{\vec{k}}^{\dag }\hat{%
\rho}_{\vec{k}}:+\dfrac{c_{1}}{2V}\sum\limits_{\vec{k}}:\hat{f}_{\vec{k}%
}^{\dag }\hat{f}_{\vec{k}}:
\end{eqnarray}
Here the following designations were used
\begin{gather}
\varepsilon _{k}=\dfrac{\hbar ^{2}k^{2}}{2M};\text{\ \ }%
c_{0}=(g_{0}+2g_{2})/3;\text{\ \ }c_{1}=(g_{2}-g_{0})/3  \notag \\
\hat{\rho}_{\vec{k}}=\sum\limits_{\vec{q},m}a_{\vec{q},m}^{\dag }a_{\vec{q}+%
\vec{k},m};\text{\ \ }\hat{\vec{f}}=(\hat{f}^{x},\hat{f}^{y},\hat{f}^{z}); \\
\hat{\vec{f}}_{\vec{k}}=\sum\limits_{q,m,n}\hat{\vec{f}}_{mn}a_{q,m}^{\dag
}a_{q+k,m}  \notag \\
\hat{f}^{x}=\left\vert
\begin{array}{ccc}
0 & 1 & 0 \\
1 & 0 & 1 \\
0 & 1 & 0%
\end{array}%
\right\vert \dfrac{1}{\sqrt{2}};\text{\ \ }\hat{f}^{y}=\left\vert
\begin{array}{ccc}
0 & -1 & 0 \\
1 & 0 & -1 \\
0 & 1 & 0%
\end{array}%
\right\vert \dfrac{i}{\sqrt{2}};\text{\ \ }\hat{f}^{z}=\left\vert
\begin{array}{ccc}
1 & 0 & 0 \\
0 & 0 & 0 \\
0 & 0 & -1%
\end{array}%
\right\vert   \notag
\end{gather}
The repeated indexes mean summation over 1,0,-1. The symbol $::$denotes the
normal ordering of the operators. The coefficient $p$ is the sum of the
linear Zeeman energy and of the Lagrangian multiplier, which is introduced
to set the total magnetization in the $z$ direction to a prescribed value.
This magnetization is conserved due to the axisymmetry of the system in a
magnetic field. q is the quadratic Zeeman effect energy, which is positive
in the case of spin $f=1$ for $^{23}$Na and $^{87}$Rb atoms. The spin-spin
interaction is of ferromagnetic-type with $c_{1} <0$ for $f=1$ $^{87}$Rb
atoms and is antiferromagnetic-type with $c_{1} >0$ for $f=1$ $^{23}$Na
atoms [10]. Taking into account that in many experimental situations the
linear Zeeman effect can be ignored and the quadratic Zeeman effect term $q$
can be manipulated experimentally, in [11] both cases of positive and
negative $q$ at $p=0$ were investigated for spin $-1$ and spin $-2$
Bose-Einstein condensates (BECs). We restrict ourselves to review some
spinor phases with spin $-1$ discussed in [11] so as to demonstrate the
relations between the Nambu-Goldstone(NG) modes of the Bogoliubov energy
spectra and the spontaneous breaking of the continuous symmetries. The
description of the excitations is presented in Refs. [10, 11] in the
number-conserving variant of the Bogoliubov theory [1]. There is no need to
introduce the chemical potential as a Lagrangian multiplier in order to
adjust the particle number to a prescribed value. The BEC takes place on a
superposition state involving the single-particle states with wave vector $%
\vec{k} =0$ and different magnetic quantum numbers
\begin{equation}
\left\vert \xi \right\rangle =\sum\limits_{m}\xi _{m}a_{0,m}^{\dag
}\left\vert vac\right\rangle ;\text{\ \ }\sum\limits_{m}\left\vert \xi
_{m}\right\vert ^{2}=1
\end{equation}
The order parameter has a vector form and consists of three components: $%
\vec{\xi } =(\xi _{1} ,\xi _{0} ,\xi _{-1} )$. The vacuum state $\left. vac
\right\rangle $ means the absence of the atoms. The ground state wave
function of the BEC-ed atoms is given by the formula (8) of Ref. [11]
\begin{equation}
\left\vert \psi _{g}\right\rangle =\dfrac{1}{\sqrt{N!}}\left(
\sum\limits_{m=-f}^{f}\xi _{m}a_{0,m}^{\dag }\right) ^{N}\left\vert
vac\right\rangle
\end{equation}
In the mean-field approximation the operators $a_{0,m}^{\dag } $, $%
a_{0,m}^{} $ are replaced by the $c-$numbers $\xi _{m} \sqrt{N_{0} } $,
where $N_{0} $ is the number of the condensed atoms. After this
substitution, the initial Hamiltonian loses its global gauge symmetry and
does not commute with the operator $\hat{N} $. The order parameters $\xi
_{m} $ are chosen so that to minimize the expectation value of the new
Hamiltonian and its ground state and to satisfy the normalization condition $%
\sum\limits_{m}\left| \xi _{m} \right| ^{2} =1$. To keep the order parameter
of each phase unchanged it is necessary to specify the combination of the
gauge transformation and spin rotations [11]. This program was carried out
in [18-21].

The initial Hamiltonian (29) in the absence of the external
magnetic field has the symmetry $U(1)\times SO(3)$ representing the global
gauge symmetry $U(1)$ and the spin-rotation symmetry $SO(3)$. The generators
of these symmetries are referred to as symmetry generators and have the form
\begin{eqnarray}
\hat{N}=\int d\vec{x}\hat{\psi}_{m}^{\dag }(x)\hat{\psi}_{m}^{{}}(x)=\sum\limits_{\vec{k},m}a_{\vec{k},m}^{\dag }a_{\vec{k},m} \notag \\
\hat{F}^{j}=\int d\vec{x}\hat{\psi}_{m}^{{}}(x)f_{mn}^{j}\hat{\psi}_{n}^{{}}(x);\text{\ \ }j=x,y,z
\end{eqnarray}
Unlike the $SO(3)$ symmetry group with three generators $\hat{F}^{x}$, $\hat{F}^{y}$ and
$\hat{F}^{z}$, the $SO(2)$ symmetry group has only one generator $\hat{F}^{z}
$ which describes the spin rotation around the $z$ axis and looks as follows:
\begin{equation}
\hat{F}^{z}=\sum\limits_{\vec{k},m}ma_{\vec{k},m}^{\dag }a_{\vec{k},m}
\end{equation}
In the presence of an external magnetic field, the symmetry of the
Hamiltonian is $U(1)\times SO(2)$. The breaking of the continuous symmetry
means the breaking of their generators. The number of the broken generators
(BG) is denoted as $N_{BG}$. There are 4 generators in the case of $%
U(1)\times SO(3)$ symmetry and two in the case of $U(1)\times SO(2)$symmetry.

The phase transition of the spinor Bose gas from the normal
state to the Bose-Einstein condensed state was introduced mathematically
into Hamiltonian (29) using the Bogoliubov displacement canonical
transformation, when the single-particle creation and annihilation operators
with the given wave vector $\vec{k} $, for example, $\vec{k} =0$, were
substituted by the macroscopically $c-$numbers describing the condensate
formation. The different superpositions of the single-particle states
determine the structure of the finally established spinor phases [11].
Nielsen and Chadha [17] formulated the theorem which establishes the
relation between the number of the Nambu-Goldstone modes, which must be
present between the amount of the collective elementary excitations, which
appear over the ground state of the system if it is formed as a result of
the spontaneous breaking of the $N_{BG} $ continuous symmetries. The number
of NG modes of the first type with linear (odd) dispersion law in the limit
of long wavelengths denoted as $N_{I} $ being accounted once, and the number
$N_{II} $ of the NG modes of the second type with quadratic (even)
dispersion law at small wave vectors, being accounted twice give rise to the
expression $N_{I} +2N_{II} $, which is equal to or greater than the number $%
N_{BG} $ of the broken symmetry generators. The theorem [17] says
\begin{equation}
N_{I}+2N_{II}\geq N_{BG}
\end{equation}
The theorem has been verified in [11] for multiple examples of the spin $-1$
and spin $-2$ Bose-Einstein condensate phases. In the case of spin $-2$
nematic phases, the special Bogoliubov modes that have linear dispersion
relation but do not belong to the NG modes were revealed. The Bogoliubov
theory of the spin $-1$ and spin $-2$ Bose-Einstein condensates (BECs) in
the presence of the quadratic Zeeman effect was developed by Uchino,
Kobayashi and Ueda [11] taking into account the Lee, Huang, Yang (LHY)
corrections to the ground state energy, pressure, sound velocity and quantum
depletion of the condensate. Many phases that can be realized experimentally
were discussed to examine their stability against the quantum fluctuations
and the quadratic Zeeman effect. The relations between the numbers of the NG
modes and of the broken symmetry generators were verified. A brief review of
the results concerning the spin $-1$ phases of [11] is presented below so as
to demonstrate, using these examples, the relations between the Bogoliubov
excitations and the Nambu-Goldstone modes.

The first example is the ferromagnetic phase with $c_{1} <0$, $q<0$ and the vector order parameter
\begin{equation}
\vec{\xi}^{F}=(1,0,0)
\end{equation}
The modes with $m=0$ and $m=-1$ are already diagonalized, whereas the mode $%
m=1$ is diagonalized by the standard Bogoliubov transformation. The
Bogoliubov spectrum is given by formulas (33) and (34) of Ref. [11]
\begin{eqnarray}
E_{\vec{k},1} &=&\sqrt{\varepsilon _{\vec{k}}(\varepsilon _{\vec{k}}+2\eta
(c_{0}+c_{1}))}; \\
E_{\vec{k},0} &=&\varepsilon _{\vec{k}}-q;\text{ }E_{\vec{k},-1}=\varepsilon
_{\vec{k}}-2c_{1}n  \notag
\end{eqnarray}
The $E_{\vec{k},1}$ mode is massless. In the absence of a magnetic field, when $q=0$, the mode $m=0$
is also massless with the quadratic dispersion law. The initial symmetry of
the Hamiltonian before the phase transition is $U(1)\times SO(3)$, whereas
the final, remaining symmetry after the process of BEC is the symmetry of
the ferromagnetic i.e. $SO(2)$. From the four initial symmetry generators $%
\hat{N}$, $\hat{F}^{x}$, $\hat{F}^{y}$ and $\hat{F}^{z}$ remains only the
generator $\hat{F}^{z}$ of the $SO(2)$ symmetry. The generators $\hat{F}^{x}$
and $\hat{F}^{y}$ were broken by the ferromagnet phase, whereas the gauge
symmetry operator $\hat{N}$ was broken by the Bogoliubov displacement
transformation. The number of the broken generators $\hat{N}$, $\hat{F}^{x}$%
, $\hat{F}^{y}$ is three, i.e., $N_{BG}=3$. In this case $N_{I}=1$, $N_{II}=1
$ and $N_{I}+2N_{II}=3$, being equal to $N_{BG}=3$. The equality $%
N_{I}+2N_{II}=N_{BG}$ takes place. In the presence of an external magnetic
field, with $q\neq 0$, the initial symmetry before the phase transition is $%
U(1)\times SO(2)$ with two generators $\hat{N}$ and $\hat{F}^{z}$, whereas
after the BEC and the ferromagnetic phase formation the remained symmetry is
$SO(2)$. Only one symmetry generator $\hat{N}$ was broken. It means $N_{BG}=1
$, $N_{I}=1$ and $N_{II}=0$. The equality $N_{I}+2N_{II}=N_{BG}$ also takes
place.

The condition $(c_{0} +c_{1} )>0$ to be hold is required for $%
m=1$ the Bogoliubov mode to be stable. It ensures the mechanical stability
of the mean-field ground state. Otherwise, the compressibility would not be
positive definite and the system would become unstable against collapse. In
the case $q>0$, $c_{1} >0$ and $(c_{0} +c_{1} )<0$ the state would undergo
the Landau instability for the $m=0$ and $m=-1$ modes with quadratic spectra
and the dynamical instability for the $m=1$ mode with a linear spectrum (36)
of Ref.[11].\newline
There are two polar phases. One with the parameters
\begin{equation}
\vec{\xi}^{P}=(0,1,0);\text{\ }q>0;\text{\ }q+2nc_{1}>0
\end{equation}
and the other with the parameters\newline
\begin{equation}
\vec{\xi}^{P^{\prime }}=\dfrac{1}{\sqrt{2}}(1,0,1);\text{\ }q<0;\text{\ }%
c_{1}>0
\end{equation}
These two polar phases have two spinor configurations which are degenerate
at $q=0$ and connect other by $U(1)\times SO(3)$ transformation. However,
for nonzero $q$ the degeneracy is lifted and they should be considered as
different phases. This is because the phase $P$ has a remaining symmetry $%
SO(2)$, whereas the phase $P^{\prime }$ is not invariant under any
continuous transformation. The number of NG modes is different in each phase
and the low-energy behavior is also different. Following formulas (40)-(42)
of [11] the density fluctuation operator $a_{kd}$ and the spin fluctuation
operators $a_{k,f_{x}}$ and $a_{k,f_{y}}$ were introduced
\begin{eqnarray}
a_{kd} &=&a_{k,0};\text{\ }a_{k,f_{x}}=\dfrac{1}{\sqrt{2}}(a_{k,1}+a_{k,-1});%
\text{\ \ }  \notag \\
a_{k,f_{y}} &=&\dfrac{i}{\sqrt{2}}(a_{k,1}-a_{k,-1});
\end{eqnarray}
Their Bogoliubov energy spectra are\newline
\begin{eqnarray}
E_{\vec{k},d} &=&\sqrt{\varepsilon _{\vec{k}}(\varepsilon _{\vec{k}}+2c_{0})}%
; \\
E_{\vec{k},f_{j}} &=&\sqrt{(\varepsilon _{\vec{k}}+q)(\varepsilon _{\vec{k}%
}+q+2nc_{1})};  \notag
\end{eqnarray}
In the presence of an
external magnetic field, the initial symmetry is $U(1)\times SO(2)$, whereas
after the BEC and the formation of the phase $P$ with $q\neq 0$ the
remaining symmetry is also $SO(2)$. Only the symmetry $U(1)$ and its
generator $\hat{N}$ were broken during the phase transition. It means we
have in this case $N_{BG}=1$, $N_{I}=1$ and $N_{II}=0$. The equality $%
N_{I}+2N_{II}=N_{BG}$ holds. Density mode is massless because the $U(1)$
gauge symmetry is spontaneously broken in the mean-field ground state, while
the transverse magnetization modes $f_{x}$ and $f_{y}$ are massive for non
zero $q$, since the rotational degeneracies about the x and y axes do not
exist being lifted by the external magnetic field. In the limit of
infinitesimal $q\rightarrow 0$ nevertheless nonzero, the transverse
magnetization modes $f_{x}$ and $f_{y}$ become massless. It occurs because
before the BEC in the absence of an external magnetic field the symmetry of
the spinor Bose gas is $U(1)\times SO(3)$, whereas after the phase
transition it can be considered as a remaining symmetry $SO(2)$. The
generators $\hat{N}$, $\hat{F}^{x}$, $\hat{F}^{y}$ were broken, whereas the
generator $\hat{F}^{z}$ remained. In this case we have $N_{BG}=3$, $N_{I}=3$
and $N_{II}=0$ the equality looks as $3=3$.

In the polar phase $P^{\prime} $ with the parameters (39) the
density and spin fluctuation operators were introduced by formulas (57)-(59)
of Ref. [11]
\begin{eqnarray}
a_{kd} &=&\dfrac{1}{\sqrt{2}}(a_{k,1}+a_{k,-1});\text{\ }a_{k,f_{x}}=a_{k,0};
\notag \\
a_{k,f_{y}} &=&\dfrac{i}{\sqrt{2}}(a_{k,1}-a_{k,-1});
\end{eqnarray}
with the Bogoliubov energy spectra described by formulas (65)-(67) [11]:
\begin{eqnarray}
E_{\vec{k},d} &=&\sqrt{\varepsilon _{\vec{k}}(\varepsilon _{\vec{k}}+2nc_{0})%
};\text{ }E_{\vec{k},f_{z}}=\sqrt{\varepsilon _{\vec{k}}(\varepsilon _{\vec{k%
}}+2nc_{1})};  \notag \\
\text{\ }E_{\vec{k},f_{x}} &=&\sqrt{(\varepsilon _{\vec{k}}-q)(\varepsilon _{%
\vec{k}}-q+2nc_{1})};\text{\ }
\end{eqnarray}
At $q<0$ in contrast to the case $q>0$ one of the
spin fluctuation mode $E_{\vec{k},f_{z}}$ becomes massless. The initial
symmetry of the system is $U(1)\times SO(2)$. It has the symmetry generators
$\hat{N}$ and $\hat{F}^{z}$. They are completely broken during the phase
transition. After the phase transition and the $P^{\prime }$ phase formation
there are not any symmetry generators. The number of the broken generator is
2 ( $N_{BG}=2$), whereas the numbers $N_{I}$ and $N_{II}$ are 2 and 0,
respectively. As in the previous cases, the equality occurs in the Nielsen
and Chadha rule. For the Bogoliubov spectra to be real the condition $q<0$, $%
c_{0}>0$ and $c_{1}>0$ must be satisfied, otherwise, the state $\vec{\xi}%
^{P^{\prime }}$ will be dynamically unstable.

Side by side with the spinor-type three-dimensional (3D) atomic
Bose-Einstein condensates in the optical traps, we will discuss also the
case of the Bose-Einstein condensation of the two-dimensional (2D)
magnetoexcitons in semiconductors [22-25]. The collective elementary
excitations under these conditions were investigated in [26-31] and will be
described in Section 11. As was shown above, the spontaneous symmetry
breaking yields Nambu-Goldstone modes, which play a crucial role in
determining low-energy behavior of various systems [5, 32-38]. Side by side
with the global gauge symmetry the local symmetry does exist.
\section{Spontaneous breaking of the local gauge symmetry and the Higgs phenomenon}
The interaction of the electrons with the electromagnetic field
can be described introducing into the Lagrangian the kinetic momentum
operators instead of canonical ones what is equivalent to introduce the
covariant derivatives $D $ instead of the differential ones $\partial $.
They are determined in Ref. [8] as
\begin{eqnarray}
\underline{x} &=&(ct,\overrightarrow{x}),\underline{\partial }=(\frac{1}{c}%
\frac{\partial }{\partial t},\overrightarrow{\nabla });  \notag \\
\underline{D} &=&\underline{\partial }-\frac{ie}{\hbar c}\underline{A};%
\underline{A}=(\varphi ,\overrightarrow{A})
\end{eqnarray}
where $\varphi $ and $\vec{A}$ are the scalar and vector potentials of the
electromagnetic field (EMF). Below we will use also the denotations of Ref.[7]
\begin{gather}
x^{\mu }=(ct,\vec{r});\text{\ x}_{\mu }=(ct,-\vec{r});\text{\ }A_{\mu
}=(\varphi ,-\vec{A});\text{\ }A^{\mu }=(\varphi ,\vec{A});  \notag \\
\partial _{\mu }=\dfrac{\partial }{\partial x^{\mu }}=\left( \dfrac{1}{c}%
\dfrac{\partial }{\partial t},\vec{\triangledown}\right) ;\text{\ }\partial
^{\mu }=\dfrac{\partial }{\partial x_{\mu }}\left( \dfrac{1}{c}\dfrac{%
\partial }{\partial t},-\vec{\triangledown}\right) ; \\
\partial _{\mu }\partial ^{\mu }=\dfrac{1}{c^{2}}\dfrac{\partial ^{2}}{%
\partial t^{2}}-\Delta;\text{\ }p^{\mu }=\left( \dfrac{E}{c},\vec{p}\right) ;%
\text{\ }p_{\mu }=\left( \dfrac{E}{c},-\vec{p}\right)   \notag
\end{gather}
The Lagrangian of the free EMF has the form [7]
\begin{equation}
L_{EMF}=-\dfrac{1}{4}F_{\mu \nu }F^{\mu \nu }
\end{equation}
being expressed through the antisymmetric tensors $F_{\mu \nu }$ and $F^{\mu
\nu }$. They are determined as four-dimensional curls of $A_{\mu }$ and $A^{\mu }$.
\begin{equation}
F_{\mu \nu }=-F_{\nu \mu }=\partial _{\mu }A_{\nu }-\partial _{\nu }A_{\mu };%
\text{\ }F^{\mu \nu }=\partial ^{\mu }A^{\nu }-\partial ^{\nu }A^{\mu }
\end{equation}
The full Lagrangian of
the electrons and EMF reads [7]
\begin{eqnarray}
L &=&\left[ \left( \partial _{\mu }+\dfrac{ie}{\hbar c}A_{\mu }\right) \phi %
\right] \left[ \left( \partial ^{\mu }-\dfrac{ie}{\hbar c}A^{\mu }\right)
\phi ^{\ast }\right] - \\
&&-m^{2}\phi ^{\ast }\phi -\lambda \left( \phi ^{\ast }\phi \right) ^{2}-%
\dfrac{1}{4}F_{\mu \nu }F^{\mu \nu }  \notag
\end{eqnarray}
As before $m^{2}$ is a parameter so
that in the case $m^{2}<0$ and in the absence of the EMF vacuum values are
determined by the formula (22).\newline
The invariance of the Lagrangian (48) under the transformation [8]
\begin{equation}
\phi ^{\prime }(\underline{x})=\phi (\underline{x})e^{i\theta (\underline{x})}
\end{equation}
in the presence of the EMF can be achieved only under the concomitant
transformation of its potential in the form [8]
\begin{equation}
\underline{A}^{\prime }(x)=A(\underline{x})+\dfrac{\hbar c}{e}\underline{\partial} \theta (\underline{x})
\end{equation}
Indeed in this case the Lagrangian (48) remains invariant [7] as follows
\begin{gather}
\left[ \left( \partial _{\mu }+\dfrac{ie}{\hbar c}A_{\mu }\right) \phi %
\right] \left[ \left( \partial ^{\mu }-\dfrac{ie}{\hbar c}A^{\mu }\right)
\phi ^{\ast }\right] =  \notag \\
=\left[ \left( \partial _{\mu }+\dfrac{ie}{\hbar c}A_{\mu }^{\prime }\right)
\phi ^{\prime }\right] \left[ \left( \partial ^{\mu }-\dfrac{ie}{\hbar c}%
A^{\prime }{}^{\mu }\right) \phi ^{\prime }{}^{\ast }\right]  \\
F_{\mu \nu }^{\prime }=F_{\mu \nu };\text{\ }F^{\prime }{}^{\mu \nu }=F^{\mu
\nu }  \notag
\end{gather}
Introducing the gauge transformation of the field
function (24) and expanding the Lagrangian in power series on the small
physical fields $\phi _{1}^{\prime }$ and $\phi _{2}^{\prime }$ we obtain
the constant, quadratic, cubic and quartic terms. The quadratic part looks
as [7]
\begin{gather}
L_{2}=-\dfrac{1}{4}F^{\mu \nu }F_{\mu \nu }+e^{2}a^{2}A_{\mu }A^{\mu }+
\notag \\
+\dfrac{1}{2}\left( \partial _{\mu }\phi _{1}^{\prime }\right) ^{2}+\dfrac{1%
}{2}\left( \partial _{\mu }\phi _{2}^{\prime }\right) ^{2}- \\
-2\lambda a^{2}\phi _{1}^{\prime }{}^{2}+\sqrt{2}eaA^{\mu }\partial _{\mu
}\phi _{2}^{\prime }  \notag
\end{gather}
The second term is proportional to $%
A_{\mu }A^{\mu }$. It indicates that the photon becomes massive. The scalar
field $\phi _{1}^{\prime }$ is also a massive one. The field $\phi
_{2}^{\prime }$ takes part in the mixed term $A^{\mu }\partial _{\mu }\phi
_{2}^{\prime }$ and can be eliminated by the supplementary gauge
transformation (50). Following Ref [7] the Lagrangian (52) can be presented
in the form
\begin{eqnarray}
L_{2} &=&-\dfrac{1}{4}F^{\mu \nu }F_{\mu \nu }+e^{2}a^{2}A_{\mu }A^{\mu }+
\notag \\
&&+\dfrac{1}{2}\left( \partial _{\mu }\phi _{1}^{\prime }\right)
^{2}-2\lambda a^{2}\phi _{1}^{\prime }{}^{2}
\end{eqnarray}
It contains two fields
only: the photon with longitudinal component and spin 1 and field $\phi
_{1}^{\prime }$ with spin 0. They are both massive. The field $\phi
_{2}^{\prime }$, which in the case of spontaneous breaking of the global
symmetry became massless forming a Goldstone boson, in this case
disappeared. The photon became massive. This phenomenon is called the Higgs
phenomenon [7].

One possible illustration of the described above effect will be
considered below following the paper by Halperin, Lee and Read [39]. They
considered the two-dimensional (2D) system of spinless electrons under the
conditions of the quantum Hall effect. Then the Hamiltonian $\hat{H} =\hat{K}
+\hat{V} $ consists of the kinetic energy operator $\hat{K}$
\begin{equation}
\widehat{K}=\frac{1}{2m_{e}}\int d^{2}\overrightarrow{r}\widehat{\psi }%
_{e}^{\dag }(\overrightarrow{r})[-i\hbar \overrightarrow{\nabla }+\frac{e}{c}%
\overrightarrow{A}(\overrightarrow{r})]^{2}\widehat{\psi }_{e}(%
\overrightarrow{r})
\end{equation}
with 2D electrons with the mass $m_{e}$ and the charge $-e$ situated in a
uniform external perpendicular magnetic field B with the vector potential $%
\vec{A}(\vec{r})$. The potential energy operator $\hat{V}$ depends on the
Coulomb interaction between the electrons. The creation and annihilation
operators $\hat{\psi}_{e}^{\dag }(\vec{r})$, $\hat{\psi}_{e}^{{}}(\vec{r})$%
obey to the Fermi statistics as was the case of Ref.[39], but we will
consider following the Ref.[40] a more general case including also the Bose
statistics
\begin{equation}
\lbrack \widehat{\psi }_{e}(\overrightarrow{r})\widehat{\psi }_{e}^{\dag }(%
\vec{r}^{\prime })\pm \widehat{\psi }_{e}^{\dag }(\vec{r}^{\prime })\widehat{%
\psi }_{e}(\overrightarrow{r})]=\delta ^{2}(\overrightarrow{r}-\vec{r}%
^{\prime })
\end{equation}
The signs $\pm $ correspond to the Fermi and Bose statistics. In Ref.[39]
the new {"}quasiparticle{"} operators $\hat{\psi}_{{}}^{\dag }(\vec{r})$, $%
\hat{\psi}(\vec{r})$ were introduced by the relations
\begin{equation}
\widehat{\psi }^{\dag }(\overrightarrow{r})=\widehat{\psi }_{e}^{\dag }(%
\overrightarrow{r})e^{-im\widehat{\omega }};\widehat{\psi }(\overrightarrow{r%
})=e^{im\widehat{\omega }}\widehat{\psi }(\overrightarrow{r})
\end{equation}
with an integer number m and with the phase operator
\begin{equation}
\hat{\omega}(\vec{r})=\int d^{2}\vec{r}^{\prime }\theta (\vec{r}-\vec{r}^{\prime })\hat{\rho}(\vec{r}^{\prime })
\end{equation}
It depends on the angle $\theta (\vec{r}-\vec{r}^{\prime })$ between the
vector $\vec{r}-\vec{r}^{\prime }$ and the in-plane axis x being determined
by the formula
\begin{equation}
\theta (\vec{r}-\vec{r}^{\prime })=\arctan \dfrac{y-y^{\prime }}{x-x^{\prime}}
\end{equation}
and by the density operator $\hat{\rho}(\vec{r}^{\prime })$
\begin{eqnarray}
\hat{\rho}(\vec{r}^{\prime})=\hat{\psi}_{e}^{\dag }(\vec{r}^{\prime })\hat{%
\psi}_{e}^{{}}(\vec{r})=\hat{\psi}_{{}}^{\dag }(\vec{r})\hat{\psi}(\vec{r}%
^{\prime })
\end{eqnarray}

These operators have the properties
\begin{eqnarray}
\hat{\omega}(\vec{r}) =\hat{\omega}^{\dag }(\vec{r});[\hat{\omega}(\vec{r}%
),\hat{\omega}^{\dag }(\vec{r}^{\prime })]=0  \notag \\
\lbrack \hat{\psi}_{e}^{{}}(\vec{r}),\hat{\rho}(\vec{r}^{\prime })] =\hat{%
\psi}_{e}^{{}}(\vec{r})\delta ^{2}(\overrightarrow{r}-\vec{r}^{\prime })
\notag \\
\lbrack \hat{\psi}_{e}^{{}}(\vec{r}),\hat{\omega}(\vec{r}^{\prime })] =%
\hat{\psi}_{e}^{{}}(\vec{r})\theta (\vec{r}-\vec{r}^{\prime }) \\
\hat{\psi}_{e}^{{}}(\vec{r})\hat{\omega}^{n}(\vec{r}^{\prime }) =(\hat{%
\omega}(\vec{r}^{\prime })+\theta (\vec{r}-\vec{r}^{\prime }))^{n}\hat{\psi}%
_{e}^{{}}(\vec{r})  \notag \\
\hat{\psi}_{e}^{{}}(\vec{r})e^{im\widehat{\omega }(\vec{r}^{\prime })}=e^{im\theta (-\vec{r}+\vec{r}^{\prime })}e^{im\widehat{\omega }(\vec{r}%
^{\prime })}\hat{\psi}_{e}^{{}}(\vec{r})  \notag \\
\hat{\psi}_{e}^{\dag }(\vec{r})e^{-im\widehat{\omega }(\vec{r}^{\prime })}
=e^{im\theta (\vec{r}-\vec{r}^{\prime })}e^{-im\widehat{\omega }(\vec{r}%
^{\prime })}\hat{\psi}_{e}^{\dag }(\vec{r})  \notag
\end{eqnarray}
It will be shown below that m is the number of
point vortices attached to each bare initial particle forming together with
it a composite particle (CP). The statistics of the CPs depends on the
statistics of the initial particles and on the number m of the attached
vortices. Finally, we will calculate the commutators of the operators $\hat{%
\psi}_{{}}^{\dag }(\vec{r})$, $\hat{\psi}_{{}}^{{}}(\vec{r})$ with the
requirement that it will be $\delta ^{2}(\vec{r}-\vec{r^{\prime }})$ as
follows:
\begin{eqnarray}
&&\lbrack \widehat{\psi }(\overrightarrow{r}),\widehat{\psi }^{\dag }(\vec{r}%
^{\prime })]_{\pm }  \notag \\
&=&e^{-im\theta (0)+im\theta (\vec{r}-\vec{r}^{\prime })}[\hat{\psi}%
_{e}^{{}}(\vec{r})\widehat{\psi }_{e}^{\dag }(\vec{r}^{\prime })\pm e^{im\pi
}\widehat{\psi }_{e}^{\dag }(\vec{r}^{\prime })\hat{\psi}_{e}^{{}}(\vec{r}%
)]\times   \notag \\
&&\times e^{-im(\widehat{\omega }(\vec{r}^{\prime })-\widehat{\omega }(\vec{r%
}))} \\
&=&\delta ^{2}(\overrightarrow{r}-\vec{r}^{\prime })  \notag
\end{eqnarray}
Here we have taken into account the relation $\theta (\vec{r}^{\prime }-\vec{%
r})-\theta (\vec{r}-\vec{r}^{\prime })=\pi $ for $\vec{r}\neq \vec{r}%
^{\prime }$. One can observe that the CPs represented by the operators $\hat{%
\psi}_{{}}^{\dag }(\vec{r})$, $\hat{\psi}_{{}}^{{}}(\vec{r})$ are composite
fermions (CFs) if the bare initial particles are fermions and the number of
vortices m is even as well as in the case when the initial particles are
bosons and the number of vortices m is odd. In the same way the CPs are
composite bosons (CBs) if the initial particles are fermions and the number
of vortices is odd, or if the initial particles are bosons and the number of
vortices m is even. \newline
\hspace{5mm} The kinetic energy operator $\hat{K}$ in terms of the operators
$\psi ^{\dag }(r)$ and $\psi (r)$ is
\begin{equation}
\hat{K}=\dfrac{\hbar ^{2}}{2m_{e}}\int d^{2}\vec{r}\psi ^{\dag }(\vec{r}%
)e^{im\hat{\omega}(\vec{r})}[-i\vec{\triangledown}+\dfrac{e}{\hbar c}\vec{A}(%
\vec{r})]^{2}e^{-}{}^{im\hat{\omega}(\vec{r})}\psi (\vec{r})
\end{equation}
It can be transformed taking into account that
\begin{eqnarray}
(-i\vec{\triangledown}+\dfrac{e}{\hbar c}\vec{A}(\vec{r}))e^{-}{}^{im\hat{%
\omega}(\vec{r})}\psi (\vec{r})  \notag \\
=e^{-}{}^{im\hat{\omega}(\vec{r})}(-i\vec{\triangledown}+\dfrac{e}{\hbar c}%
\vec{A}(\vec{r})-m\vec{\triangledown}\hat{\omega}(\vec{r}))\psi (\vec{r})
\end{eqnarray}
what leads to the formula
\begin{equation}
\hat{K}=\dfrac{\hbar ^{2}}{2m_{e}}\int d^{2}\vec{r}\psi ^{\dag }(\vec{r})[-i%
\vec{\triangledown}+\dfrac{e}{\hbar c}\vec{A}(\vec{r})-m\vec{\triangledown}%
\hat{\omega}(\vec{r})]^{2}\psi (\vec{r})
\end{equation}
It contains a supplementary vector potential $\hat{\vec{a}}(\vec{r})$ named as statistical Chern-Simons gauge
potential [41] determined as
\begin{equation}
\hat{\vec{a}}(\vec{r})=-\dfrac{m\hbar c}{e}\vec{\triangledown}\hat{\omega}(%
\vec{r})=-\dfrac{m\hbar c}{e}\int d^{2}\vec{r}^{\prime }\vec{\triangledown}%
_{r}\theta (\vec{r}-\vec{r}^{\prime })\hat{\rho}(\vec{r}^{\prime })
\end{equation}
Its calculation needs a special
precaution as was pointed by Jackiw and Pi [41] because $\theta (\vec{r}-%
\vec{r}^{\prime })$ is a multivalued function. They cautioned against the
moving of $\vec{\triangledown}$ with respect to $\vec{r}$ out of the
integral $\int d^{2}\vec{r}^{\prime }\theta (\vec{r}-\vec{r}^{\prime })\hat{%
\rho}(\vec{r}^{\prime })$, because in general it is not correct. The
integration cannot be interchanged with the differentiation. The reason for
this is that the function $\theta (\vec{r}-\vec{r}^{\prime })$ is
multivalued and the integration of $\theta (\vec{r}-\vec{r}^{\prime })$ over
the two-dimensional $\vec{r}^{\prime }$ plane requires specifying the cut in
the space $\vec{r}^{\prime }$, which begins at the point $\vec{r}$. The
range of the $\vec{r}^{\prime }$ integration depends on $\vec{r}$ and moving
the $\vec{r}$ derivative outside of the $\vec{r}^{\prime }$ integral gives
an additional contribution. To avoid these complications the derivative $%
\vec{\triangledown}\theta (\vec{r}-\vec{r}^{\prime })$ is introduced into
the integrand for the very beginning in the form
\begin{equation}
\vec{\triangledown}\theta (\vec{r}-\vec{r}^{\prime })=-curl\ln |\vec{r}-\vec{%
r}^{\prime }|
\end{equation}
As it was shown in Ref. [41] the curl of a scalar S in the 2D space is a
vector and the curl of the vector $\vec{a}$ is a scalar as follows
\begin{eqnarray}
(curlS)^{i} &=&\in ^{ij}\partial _{j}S;\text{\ }curl\vec{a}=\in
^{ij}\partial _{i}a_{j};\text{\ }i,j=1,2  \notag \\
&\in &^{12}=-\in ^{21};\in ^{11}=\in ^{22}=0
\end{eqnarray}
The statistical gauge
vector potential $\hat{\vec{a}}(\vec{r})$ can be transcibed
\begin{equation}
\hat{\vec{a}}(\vec{r})=\dfrac{m\hbar c}{e}\int d^{2}\vec{r}^{\prime
}curl\ln |\vec{r}-\vec{r}^{\prime }|\hat{\rho}(\vec{r}^{\prime })
\end{equation}
what leads to the statistical gauge magnetic field $\hat{b}(\vec{r})$
\begin{gather}
\hat{b}(\vec{r})=curl\hat{\vec{a}}(\vec{r})=\dfrac{m\hbar c}{e}\int d^{2}%
\vec{r}^{\prime }\in ^{ij}\in ^{jk}\partial _{j}\partial _{k}\ln |\vec{r}-%
\vec{r}^{\prime }|\hat{\rho}(\vec{r}^{\prime })=  \notag \\
=-\dfrac{m\hbar c}{e}\int d^{2}\vec{r}^{\prime }\Delta_{r}\ln |\vec{r}-\vec{r}%
^{\prime }|\hat{\rho}(\vec{r}^{\prime })
\end{gather}
Taking into account the equality
\begin{equation}
\Delta \ln \vec{r}=2\pi \delta ^{2}(\vec{r})
\end{equation}
we obtain\newline
\begin{equation}
\hat{b}(\vec{r})=-\dfrac{2\pi m\hbar c}{e}\hat{\rho}(\vec{r})
\end{equation}
Substituting the density operator $\hat{\rho}(\vec{r})$ by its
mean value $n_{e}=\dfrac{\nu }{2\pi l^{2}}$ with the fractional integer
filling factor $\nu $ equal to $\nu =1/m$ with $m\geq 1$, and taking into
account the magnetic length $l^{2}=\dfrac{\hbar c}{eB}$ determined by the
external magnetic field B we will find the average value $\bar{b}$ and
equality
\begin{equation}
B+\bar{b}=0
\end{equation}
what means that the resulting magnetic field is exactly zero.
In this approximation the set of CPs does exist in zero magnetic field. If
they are fermions their ground state will be a filled Fermi sea with the
Fermi wave vector determined by the magnetic length. If they are bosons,
they will undergo the BEC.

In the Section 7 we will discuss the collective elementary
excitations above the ground state in the case of CBs on the base of
Ginzburg-Landau theory. In Ref.[42] it was shown that applying the
mean-field theory one must integrate out the short-distance fluctuations of
the $\psi (r)$ field to obtain an effective action which describe the
physics at distance scales larger than the magnetic length. It is supposed
that the effective action is of the same form as the microscopic action, but
with renormalized stiffness constant, bare mass and the effective
interaction strength.
\section{Quasi-Nambu-Goldstone Modes in the Bose-Einstein Condensates}
The Goldstone theorem guarantees that the NG modes do not acquire
mass at any order of quantum corrections. Nevertheless, sometimes
soft modes appear, which are massless in the zeroth order but
become massive due to quantum corrections. They were introduced
by Weinberg [5], who showed that these modes emerge if the symmetry
of an effective potential of the zeroth order is higher than
that of the gauge symmetry and the idea was invoked to account
for the emergence of low-mass particles in relativistic physics.
Following [32] now these modes are referred to as quasi-Nambu-Goldstone
modes, in spite of the fact that their initial name introduced
by Weinberg was pseudo-modes instead of quasi-modes. Georgi and
Pais [33] demonstrated that the quasi-NG modes also occur in
cases in which the symmetry of the ground state is higher than
that of the Hamiltonian [32]. This type of the quasi-Nambu-Goldstone
modes is believed to appear, for example, in the weak-coupled
limit of A phase of $^{3}$He [37, 38].

The authors of [32] underlined that the spinor BEC are ideal
systems to study the physics of the quasi-NG modes, because these
systems have a great experimental manipulability and well established
microscopic Hamiltonian. It was shown in [32] that the quasi-NG
modes appear in a spin-2 nematic phase. In the nematic condensate,
three phases, each of which has a different symmetry, are energetically
degenerate to the zeroth order [36] and the zeroth order solution
has a rotational symmetry
$SO(5)$, whereas the Hamiltonian of the spin-2 condensate has a rotational
symmetry
$SO(3)$. By applying the Bogoliubov theory of the BEC under the assumption
that the
$\vec{k} =0$ components of the field operators are macroscopically occupied,
it was shown that the order parameter of the nematic phase has
an additional parameter independent on the rotational symmetry.

The ground state symmetry of the nematic phase at the zeroth
order approximation is broken by quantum corrections, thereby
making the quasi-NG modes massive. The breaking of the
$SO(5)$ symmetry occurs. The number n of the quasi-NG modes was determined
by Georgi and Pais [33] in the form of the theorem. It was explained
and represented in [32] as follows:
\begin{equation}
n=\dim (\tilde{M} )-\dim (M)
\end{equation}
where $\tilde{M} $ is the surface on which the effective potential assumes its
minimal values to the zeroth order and
$\dim (\tilde{M} )$ is the dimension of this surface. The dimension
$\dim (M)$ determines the number of the NG modes. This implies that
$M$ is a submanifold of $\tilde{M} $ and n is the dimension of the complementary space of
$M$ inside $\tilde{M} $ [32].

In the case considered by Goldstone, the dimension of the ring
is 1 and the number of the NG modes is 1. This leads to the absence
of the quasi-NG modes
$(n=0)$. Returning to the case of 2D magnetoexcitons in the BEC state
with small but nonzero wave vector $\vec{k} $
$(\vec{k} \neq 0)$ described by Hamiltonian (16) of [30], one should remember that
both continuous symmetries existing in the initial form (10)
[30] were lost. It happened due the presence of the term
$\tilde{\eta } (d_{\vec{k} }^{\dag } +d_{\vec{k} } )$ in the frame of the Bogoliubov theory of quasiaverages. Nevertheless,
the energy of the ground state as well as the self-energy parts
$\Sigma _{ij} (P,\omega )$, which determine the energy spectrum of the collective elementary
excitations depend only on the modulus of the wave vector
$\vec{k} $ and do not depend at all on its direction. All these expressions
have a rotational symmetry
$SO(2)$ in spite of the fact that Hamiltonian (16) of [30] has lost
it. We believe that the condition described by Georgi and Pais
[33] favoring the emergence of the quasi-NG modes. We are explaining
the existence of the gapped, massive exciton-type branches of
the collective elementary excitations obtained in our calculations
just by these considerations. These questions will be discussed
in Section 11.
\section{Ginzburg-Landau theory for the fractional quantum Hall effect}
In this section we will follow the collective monograph [43]
dedicated to the fractional quantum Hall effect (FQHE), the clear and
transparent candidatus scientiarum thesis by Enger [44] and many other
papers cited below. The Landau theory of the second order phase transition
[45] is based on the introduction of the order parameter, $\phi (rt)$
assuming that the free energy is a regular function of $\phi $ at least near
the critical point. In the case of superconductors and superfluids the role
of the order parameter is played by the condensate wave functions. The
theory of superconductors was elaborated by Ginzburg and Landau [46] whereas
for liquid helium by Ginzburg and Pitaevskii [47]. The microscopical
foundations in the latter case were proposed by Pitaevskii [48] and by Gross
[49] and can be found in the monograph by Nozieres and Pines [50]. The
microscopical theory of superfluidity was firstly proposed by Bogoliubov as
the model of weakly interacting Bose gas [51]. The density of the Helmholz
free energy $f(r)$ expanded on the small order parameter $\phi $ has the form
\begin{equation}
f(r)=f_{0}+\alpha \left\vert \phi \right\vert ^{2}+\dfrac{\beta }{2}%
\left\vert \phi \right\vert ^{4}+\dfrac{\hbar ^{2}}{2m}\left\vert
\triangledown \phi \right\vert ^{2}
\end{equation}
In the case of superconductor it is necessary to include the effect of the
applied electromagnetic field which can be done by substituting the
canonical momentum $\hat{p} =-i\hbar \triangledown $ by the kinetic momentum
\begin{equation}
\vec{p}-\dfrac{q}{c}\vec{A}(r)
\end{equation}
where $\vec{A} $ is the vector potential, q is the charge of the Cooper
pair, $q=-2e$.\newline
The density of the Gibbs free energy including also the density of the
magnetic field energy looks as
\begin{eqnarray}
g(r) &=&f_{0}+\alpha \left\vert \phi \right\vert ^{2}+\dfrac{\beta }{2}%
\left\vert \phi \right\vert ^{4}+ \notag \\
&&+\dfrac{1}{2m}\left\vert \left( -i\hbar \triangledown +\dfrac{2e}{c}\vec{A}%
\right) \phi \right\vert ^{2}+\dfrac{B^{2}}{2\mu _{0}}
\end{eqnarray}
where $\vec{B}=rot\vec{A}$. Minimizing the total Gibbs energy $G=\int g(r)dr
$ with respect to $\phi $ and $\vec{A}$ gives
\begin{gather}
\dfrac{1}{2m}\left( -i\hbar \triangledown +\dfrac{2e}{c}\vec{A}\right)
^{2}\phi +\alpha \phi +\beta \left\vert \phi \right\vert ^{2}\phi =0 \notag \\
\dfrac{1}{\mu _{0}}\vec{\triangledown}\times \vec{B}=\dfrac{ie\hbar }{m}%
\left( \phi ^{\ast }\triangledown \phi -\phi \triangledown \phi ^{\ast
}\right) -\dfrac{4e^{2}}{mc^{2}}\left\vert \phi \right\vert ^{2}\vec{A}
\end{gather}
This is the Ginzburg-Landau equations, where $\mu _{0} $ is the magnetic
permeability.\newline
The Ginzburg-Pitaevskii-Gross equation for the Bose-Einstein condensate wave
function $\phi (r,t)$ is
\begin{equation}
i\hbar \dfrac{\partial \phi (r,t)}{\partial t}=-\dfrac{\hbar ^{2}}{2m}?\phi
(r,t)+\lambda \left\vert \phi (r,t)\right\vert ^{2}\phi (r,t)
\end{equation}
Separating the space and time parts $\phi (r,t)=e^{-i\mu t}\phi (r)$, and
choosing the chemical potential $\mu =\lambda \rho _{0}$, one can transform
(78) into the equation
\begin{equation}
-\dfrac{\hbar ^{2}}{2m}\Delta \phi (r)+\lambda \left( \left\vert \phi
(r)\right\vert ^{2}-\rho _{0}\right) \phi (r)=0
\end{equation}
which is known as Gross-Pitaevskii equation or non-linear Schrodinger equation. As
was mentioned in [51, 52] the GL theory is needed also for the FQHE to
better understand this phenomenon.

The FQHE also is a remarkable example of the quantum effects
observable on a macroscopic level similarly as superconductivity and
superfluidity. All these phenomena have a ground state with non-zero density
of particles and in all three cases there are quasiparticle excitations in
the form of vortices. But there are some aspects of the FQHE, which are not
present in the GL theories of superconductors and superfluids. First of them
there is a gap in the spectrum of the collective elementary excitations,
which leads to the incompressibility of the FQHE systems. The second
important difference is related with the properties of the vortices in the
FQHE case. They play the role of the single-particle excitations and have
finite creation energy, as opposed to the vortices in the superfluid He-II
with an extensive creation energy of the vortex proportional to $\ln (R/a)$,
where R is the radius of the system and a is the vortex core.\newline
\hspace{5mm} In addition the FQHE vortices have fractional charges [52]. In
numerous papers some variants of the G-L theory for the FQHE were proposed
starting with the Lagrangian of the system containing the supplementary term
known as Chern-Simons term. It describes the gauge vector potential
generated by the vortices; which in their turn are induced by the flux
quanta created by the external magnetic field B. Instead of Gibbs free
energy the action of the system is studied. \newline
Girvin [52], and Girvin and MacDonald [53] for the first time
proposed a phenomenological variant of the GL theory writing the action S in
the form
\begin{eqnarray}
S=\int d^{2}r\{\left\vert \left( -i\hbar \triangledown +\dfrac{e}{c}%
A_{1}(r)\psi (r)\right) \right\vert ^{2}+  \notag \\
+i(\psi ^{\ast }(r)\psi (r)-n_{0})\phi (r)- \\
\_\dfrac{i\theta }{8\pi ^{2}}(\phi \triangledown \times \vec{A}_{1}+\vec{A}%
_{1}\times \triangledown \phi )\}  \notag
\end{eqnarray}
where
\begin{equation}
\vec{A}_{1}=\vec{A}+\vec{a};\text{\ }\vec{B}=rot\vec{A};
\end{equation}
is an effective summary vector potential composed from the
physical external vector potential $\vec{A}$ generating the magnetic field
B, and from a gauge vector potential $\vec{a}$ created by the vortices. The
effective field $A_{1}$ represents the frustration arising in the system,
when the density of the particles $\rho (r)=\left\vert \psi (r)\right\vert
^{2}$ deviates away from the quantized Lauglin's density $n_{0}$ [42], which
determines the fractional filling factor $\nu =1/m$ with m integer. The
density $n_{0}$ is named the flux density being determined by the magnetic
field B through the magnetic length $l$ in the form $n_{0}=1/m2\pi l^{2}$,
where $l^{2}=\hbar c/eB$. The equation of motion for vector $\vec{A}_{1}$ in
a static case is:
\begin{equation}
\theta \triangledown \times \vec{A}_{1}=(\psi ^{\ast }\psi -n_{0});\text{\ }%
\theta =2\pi /m
\end{equation}
The proposed phenomenological G-L theory allows us to
understand that the creation energy of a single vortex is finite and that
the vortex has a fractional charge. The difference between the FQHE and
ordinary superfluidity was explained by the strong phase fluctuations
induced by the frustration. Zhang, Hansen and Kivelson [42] derived their
field-theory model starting from the microscopic Hamiltonian. They
constructed the G-L theory in a way similar to Girvin but contrary to Girvin
in their approach the Chern-Simons term contains only the gauge field $a(r)$
[42]. As in the previous papers [53] it was confirmed that the disturbances
of the localized density moving the system away from the good filling
fractions lead to creation of single-particle excitations. These
quasiparticle and quasihole excitations have the form of vortices with
static nonuniform finite-energy solutions. Side by side with the
single-particle excitations in the Ref.[42] the collective elementary
excitations were discussed. For this end the Lagrangian was expanded up to
terms quadratic in $\delta \phi $ and $\delta a$ about the constant
solutions corresponding to vacuum expectation values. The fluctuating values
$\delta \phi $ and $\delta a$ were represented in the form of plane waves
with vector q. The dispersion relation was found in the form [42]
\begin{equation}
\omega ^{2}(q)=(e\kappa B)^{2}+\dfrac{1}{4}\kappa q^{2}(\kappa
q^{2}+8\lambda n_{0})
\end{equation}
It has a gap in the point $q=0$ proportional to the external magnetic field
B. For negativ $\lambda $, but for sufficiently small parameter $\left|
\lambda \right| /\kappa $ the dispersion curve has a roton-type behavior
with the same shape as was derived by Girvin, MacDonald and Platzman [54].
The GL theory proposed by [42] describes the incompressibility, fractional
charge and fractional statistics of the quasiparticles. But being a
coarse-grained version of the FQHE it makes errors on the magnetic length
scale. It treats the gauge field with a mean-field approximation, and
reproduces correctly the long-wavelength effects of the quantum Hall systems
excluding such details as the description of the vortex core. The idea that
the long-wavelength effects of the physical magnetic field are canceled by
the gauge field was also suggested by Laughlin [55] and in Ref.[56].
\section{Point vortices under the conditions of FQHE}
Because the vortices play an important role in the
understanding of the FQHE we will provide here more information on this
subject. The presentation below beginning with classical hydrodinamics and
proceeding to the quantum vortices is given following the paper [57] and
Enger [44] and Myklebust theses [58]. An ideal fluid without viscosity is
described in classical hydrodynamics by the continuity equation.
\begin{equation}
\dfrac{\partial \rho }{\partial t}+\vec{\triangledown}(\rho v)=0
\end{equation}
and Euler's equation
\begin{equation}
\dfrac{\partial v}{\partial t}+(v\cdot \vec{\triangledown}\text{)}v=-\dfrac{%
\vec{\triangledown}p}{\rho }
\end{equation}
where $\rho ,\text{\ } p$ and $\vec{v} $ are the density, pressure and
velocity field correspondingly in each point of the liquid. The vorticity is
defined in 3D hydrodynamics as $\vec{\omega } =\vec{\triangledown } \times
\text{v} $. If the liquid is not only ideal but also isentropic with
constant entropy along it, then the vorticity $\vec{\omega } $ obeys to a
supplementary continuity equation. The flow is irrotational with a potential
flow if $\vec{\omega } =0$ at all points of the fluid. In this case one can
introduce the velocity potential $\phi $
\begin{equation}
v=\vec{\triangledown}\phi ;\ \vec{\omega}=\vec{\triangledown}\times v=0
\end{equation}
In physical fluids the vorticity is localized in small areas. Outside the
vortices most of fluid is irrotational. In a 3D liquid the vortex is a tube
with the strength $\kappa $ defined as
\begin{equation}
\kappa =\int \vec{\omega}d\vec{\sigma}=\oint vd\vec{l}
\end{equation}
The Helmholtz theorem (also known as Kelvin's circulation theorem) says that
in the absence of rotational external forces a fluid that is initially
irrotational remains irrotational all the time. In case of 2D fluid the
notion of point vortex with zero area is introduced. The velocity field
generating such a vortex may be represented by the expressions
\begin{gather}
v=\dfrac{\kappa }{2\pi r}\vec{e}_{Q}=\dfrac{\kappa }{2\pi }\left( -\vec{i}%
\dfrac{y}{r^{2}}+\vec{j}\dfrac{x}{r^{2}}\right) ;  \notag \\
\vec{e}_{\theta }=\vec{j}\cos \theta -\vec{i}\sin \theta ;\text{\ }\vec{e}%
_{r}=\vec{i}\cos \theta +\vec{j}\sin \theta ; \\
\vec{\triangledown}=\vec{i}\dfrac{\partial }{\partial x}+\vec{j}\dfrac{%
\partial }{\partial y}=\dfrac{\partial }{\partial r}\vec{e}_{r}+\dfrac{1}{r}%
\dfrac{\partial }{\partial \theta }\vec{e}_{\theta }  \notag
\end{gather}
Here $\kappa $ is the vortex strength, whereas the unit vectors $\vec{i} ,%
\text{\ } \vec{j} ,\text{\ } \vec{e} _{r} $ and $\vec{e} _{\theta } $
corresponds to rectangular and polar 2D coordinates. Following Ref [41] we
must take into account the definition of the curl in the 2D space, namely
that the curl of the vector is a scalar and the curl of the scalar is a
vector as follows
\begin{equation}
\omega =Curlv=\vec{\triangledown}\times v=\varepsilon ^{ij}\partial
_{i}v_{j};\text{\ }\left( CurlS\right) ^{i}=\varepsilon ^{ij}\partial _{j}S
\end{equation}
where $\varepsilon ^{ij}$ is an antisymmetric tensor with the properties $%
\varepsilon ^{12}=-\varepsilon ^{21}=1;\text{\ }\varepsilon
^{11}=\varepsilon ^{22}=0$.\newline
These rules lead to the vorticity of the point vortex with the velocity
field (88)
\begin{equation}
\omega (r)=Curlv=\dfrac{\kappa }{2\pi }\Delta \ln r=\kappa \delta ^{(2)}(%
\vec{r})
\end{equation}
The velocity field created by a point vortex has a singularity. It is
irrotational or potential almost in all space except of the origin in the
point $r=0$. By this reason the vortex area is zero. Nevertheless the
summary vorticity due the singularity (90) is finite. In the same way the
circulation of the vortex is also finite as follows
\begin{equation}
\int \omega d^{2}\vec{r}=\lim\limits_{r\rightarrow \infty }\oint \dfrac{%
\kappa }{2\pi r}\vec{e}_{\theta }d\vec{l}=\kappa ;\text{\ }d\vec{l}=rd\theta
\vec{e}_{\theta }
\end{equation}
A fluid containing a point vortex will have potential flow almost
everywhere. A point vortex in an incompressible liquid has energy
\begin{equation}
\int \dfrac{mv^{2}}{2}d^{2}\vec{r}=\dfrac{m\kappa ^{2}}{2\pi }\ln \dfrac{R}{%
a}
\end{equation}
where R is the length scale of the whole system and a is the core radius. A
classical system of N point vortices in an incompressible liquid has the
kinetic energy associated with each vortex and the interaction energy
between them. This interaction does not come from an electric charge of the
vortices because they are neutral. For two vortices with guiding centers $%
\vec{R} _{1} $ and $\vec{R} _{2} $ it is useful to define a guiding center
of a pair $\vec{R} _{gc} $ and its relative coordinate $\vec{R} _{rel} $ in
the form
\begin{equation}
\vec{R}_{gc}=\vec{R}_{1}+\vec{R}_{2};\vec{R}_{rel}=\vec{R}_{1}-\vec{R}_{2}
\end{equation}
The equations of motion for a pair of vortices with equal strengths $\kappa
_{1} =\kappa _{2} =\kappa $ are
\begin{equation}
\dot{X}_{gc}=\dot{Y}_{gc}=0;\dot{X}_{rel}=-\dfrac{\kappa Y_{rel}}{\pi
R_{rel}^{2}};\dot{Y}_{rel}=\dfrac{\kappa }{\pi }\dfrac{X_{rel}}{R_{rel}^{2}}
\end{equation}
These equations describe a circular motion around a fixed point named as
stationary guiding center with an angular velocity $?$ depending on the
constant separation distance of the vortices $\left| \vec{R} _{rel} \right| $
as follows
\begin{equation}
\Omega =\dfrac{\kappa }{\pi \left\vert \vec{R}_{rel}\right\vert ^{2}}
\end{equation}
For a pair of vortices with opposite vorticities $\kappa =\kappa _{1}
=-\kappa _{2} $ i.e. for a vortex-antivortex pair the equations of motion are
\begin{equation}
\dot{X}_{gc}=\dfrac{\kappa }{\pi }\dfrac{Y_{rel}}{\left\vert \vec{R}%
_{rel}\right\vert ^{2}};\dot{Y}_{gc}=-\dfrac{\kappa }{\pi }\dfrac{X_{rel}}{%
\left\vert \vec{R}_{rel}\right\vert ^{2}};\dot{X}_{rel}=\dot{Y}_{rel}=0
\end{equation}
The vortices will not move relative each other, but will follow a straight
line perpendicular to the vector $\vec{R} _{rel} $ connecting the vortices
[44].\newline
This picture is exactly the same as the structure of a 2D magnetoexciton
moving with wave vector $\vec{k} $ perpendicular to the vector $\vec{d} $
connecting the electron and hole in the pair with a constant distance $%
d=kl^{2} $ at a given $\vec{k} $. \newline
One can remember that the existence of the quantum vortices was suggested
for the first time by Onsager [59], who proposed that the circulation in the
superfluid He-II is quantized with the quantum of circulation $h/m$. The
quantum vortices in the He-II were discussed by Feynman [60], whereas a
quantized line was observed by Vinen [61]. The quantization of the vorticity
in the He-II can be explained in the frame of GL theory. The velocity field
of a superfluid described by the wave function
\begin{equation}
\phi =\sqrt{\rho }e^{iS}
\end{equation}
can be written as
\begin{equation}
v=\dfrac{\hbar }{m}\vec{\triangledown}S
\end{equation}
The circulation around a close path C becomes
\begin{equation}
\kappa =\oint vd\vec{l}=\dfrac{\hbar }{m}\oint \vec{\triangledown}Sd\vec{l}%
=\dfrac{\hbar }{m}\delta S
\end{equation}
$\delta S$ is the change in the phase of the wave function, as one moves
around the close path C. But the wave function must be single valued. By
this reason $\delta S$ must be integer multiple of $2\pi $. It means that
\begin{equation}
\kappa =\dfrac{\hbar }{m}2\pi s,\ s=0,\pm 1,\pm 2...
\end{equation}
The vorticity of the quantum vortex has discrete values with the quanum $h/m$. This definition of vorticity differs from the classical hydrodynamics $%
\vec{\omega } =\vec{\triangledown } \times \text{v} $. The only rotational
invariant wave function having the property (100) being written in polar
coordinated has $S=s\theta $
\begin{equation}
\phi (\vec{r})=f(r)e^{is\theta }
\end{equation}
It produces the same velocity field as the classical point vortex
\begin{equation}
v=\dfrac{\hbar s}{mr}\vec{e}_{\theta }=\dfrac{\kappa }{2\pi r}\vec{e}%
_{\theta }
\end{equation}
The kinetic energy
\begin{equation}
E=\int \dfrac{1}{2}mv^{2}d^{2}r=\dfrac{\hbar ^{2}\pi }{m}s^{2}\ln \left(
R/\xi \right)
\end{equation}
is now expressed through the coherence length $\xi $ instead of the core
radius $a$. The cutoff at $\xi $ is used to avoid the logarithmic divergence
neared the vortex core. \newline
Inserting the vortex function (101) into the Ginzburg-Pitaevskii-Gross
equation, Myklebust [44, 58] found the following equation for the function $f(r)$
\begin{equation}
\dfrac{d^{2}f}{dr^{2}}+\dfrac{1}{r}\dfrac{\partial f}{\partial r}+\left( 2-%
\dfrac{s^{2}}{r^{2}}\right) f-2f^{3}=0.
\end{equation}
It depends only on $s^{2} $. Contrary to the He-II, the Bose-Einstein
condensate in superconductors is formed by the Cooper pairs with the charge $%
q=-2e$ instead of the neutral atoms. The type II superconductors allow the
magnetic field to penetrate in metals forming quantized vortices, while in
the type-I superconductors the magnetic field cannot penetrate [62]. The
quantized vortices exist in the form of filaments named Abrikosov's lines.
They have a mixed electron and electromagnetic field origin and were
described for the first time by Abrikosov [63] on the base of the G-L theory
with nonzero electromagnetic field $\vec{A} $. It was shown that the
magnetic flux through the vortex tube is quantized with the flux quantum $%
\phi _{0} $
\begin{equation}
\int \vec{B}d\vec{\sigma}=\oint \vec{A}d\vec{l}=n\phi _{0};\text{\ }\phi
_{0}=\dfrac{2\pi \hbar c}{\left\vert q\right\vert }
\end{equation}
The total energy per a unit length of the vortex tube is finite and equals to
\begin{equation}
E=\left( \dfrac{\phi _{0}}{4\pi \lambda }\right) ^{2}\ln \dfrac{\lambda }{%
\xi };\text{\ \ }\lambda >\xi
\end{equation}
where $\lambda $ is the penetration length of the magnetic field into the
II-type superconductors as was introduced by F. London and H. London [64]
and $\xi $ is the correlation length between the electrons in the Cooper
pair. Girvin [52] suggested that contribution of the electromagnetic field
in the resultant current density $\vec{j} (\vec{r} )$ determined in the case
of FQHE as
\begin{eqnarray}
\vec{j}(\vec{r}) &=&\dfrac{1}{2}\left\{ \psi ^{\ast }(r)(-i\hbar \vec{%
\triangledown})\psi (r)+\psi (r)(i\hbar \vec{\triangledown})\psi ^{\ast
}(r)\right\}   \notag \\
&&+\dfrac{e}{c}\vec{A}\psi ^{\ast }(r)\psi (r)
\end{eqnarray}
reorganizes the point vortex state in such a way that its resultant
circulation at great distance $r\rightarrow \infty $ will be zero
\begin{equation}
\oint \vec{j}(r)d\vec{l}=0
\end{equation}
It is possibly only for the condition when the magnetic flux through the
vortex surface is quantized in the form
\begin{equation}
\int rot\vec{A}d^{2}\vec{r}=\oint \vec{A}d\vec{l}=-m\phi _{0};\text{\ \ }%
\phi _{0}=\dfrac{2\pi \hbar c}{\left\vert e\right\vert }
\end{equation}
This value being multiplied by $n_{0} \left| e\right| /c$ compensates
exactly the circulation arising from the electron part of the current density
\begin{eqnarray}
\oint \dfrac{1}{2}\left\{ \psi ^{\ast }(r)(-i\hbar \vec{\triangledown}%
)\psi (r)+\psi (r)(i\hbar \vec{\triangledown})\psi ^{\ast }(r)\right\} d\vec{%
l}  \notag \\
=2\pi \hbar mn_{0}
\end{eqnarray}
because the wave function $\psi (r)$ has the form
\begin{equation}
\psi (r)=\sqrt{n_{0}}f(r)e^{im\theta };\text{\ }f(r)\rightarrow 1;\text{\ }%
r\rightarrow \infty
\end{equation}
The number of magnetic flux quanta $-m$ must be opposite to the magnetic
quantum number of the electron wave function. The creation energy of such
point vortex is finite and no extensive as in the case of a pure electron
vortex. As was mentioned by Girvin and MavDonald [53] the isolated vortex
cost only a finite energy. They can be excited thermally by one. Earlier it
was necessary to create a pair vortex-antivortex with finite creation energy
for a pair as a whole, but with extensive energy for each of them. Only in
the last case the Kosterlitz-Thoulless phase transition was possibly being
related with the unbinding of the vortices in the pairs.
\section{Gauge transformations and statistical gauge field}
Girvin and MacDonald [53] revealed a hidden symmetry of the
Laughlin's [65] ground state wave function describing the FQHE of the 2D
one-component electron gas (OCEG). This wave function is
\begin{equation}
\psi (z_{1},...,z_{N})=\prod\limits_{i<j}\left( z_{i}-z_{j}\right) ^{m}\exp
\left[ -\dfrac{1}{4}\sum\limits_{k}\left\vert z_{k}\right\vert ^{2}\right]
\end{equation}
The filling factor of the lowest Landau level (LLL) is a fractional integer $%
\nu =1/m$, with integer $m>1$. $z_{k} =x_{k} +iy_{k} $are the complex
coordinates of the particles in symmetric gauge. With respect to the
interchanging of any two particles the wave function (112) is anti-symmetric
at odd values of m and symmetric at even values, describing the fermions and
bosons, correspondingly. But changing the phase of the wave function (111)
using a singular gauge transformation
\begin{eqnarray}
\psi _{new}(z_{1},...,z_{N}) =\exp \left[ -im\sum\limits_{i<j}\arg \left(
z_{i}-z_{j}\right) \right] \times   \notag \\
\times \psi (z_{1},...,z_{N}) =\prod\limits_{i<j}\left\vert
z_{i}-z_{j}\right\vert ^{m}\exp \left[ -\dfrac{1}{4}\sum\limits_{k}\left%
\vert z_{k}\right\vert ^{2}\right]
\end{eqnarray}
we have obtained a bosonic type wave function at
any integer values of $m>1$. The off-diagonal matrix elements of the density
matrix $\rho (z,z^{\prime })$ calculated with the function (111) are
short-ranged with a characteristic scale given by the magnetic length, while
those calculated with the wave function (112) $\tilde{\rho}(z,z^{\prime })$
have a slowly decreasing behavior with a power law $\left\vert z-z^{\prime
}\right\vert ^{-m/2}$. The singular gauge density matrix $\tilde{\rho}%
(z,z^{\prime })$ has an off-diagonal long-range order (ODLRO). The physical
origin of this difference is related to the presence of the vortices induced
around each particle under the influence of the magnetic flux quanta, as was
explained by Stormer [66].

The presence of the vortices can be demonstrated using more
simple example proposed by Enger [44] with a wave function $\psi (z)$ of two
particles depending only on their relative coordinate $z$. It is supposed
that $\psi (z)$ obeys to any statistics and after the particle interchanging
it becomes
\begin{gather}
\psi (e^{i\pi }z)=e^{i\theta }\psi (z);\text{\ \ \ \ }%
\begin{array}{c}
\theta =\pi (2n+1)\text{\ \ \ for\ fermions} \\
\theta =2\pi n\text{\ \ \ \ \ \ \ \ \ \ \ \ \ \ for\ bosons}%
\end{array}
\notag \\
n=0,\pm 1,\pm 2...
\end{gather}
A gauge transformation $e^{i\eta (z)} $ transforms the wave function $\psi
(z)$ into another bosonic type wave function
\begin{equation}
e^{i\eta (z)}\psi (z)=\phi (z)=\phi (e^{i\pi }z)
\end{equation}
To satisfy this requirement and the equalities
\begin{equation}
e^{i\eta \left( e^{i\pi }z\right) }\psi \left( e^{i\pi }z\right) =e^{i\eta
(z)}\psi (z)=e^{i\eta \left( e^{i\pi }z\right) }e^{i\theta }\psi (z)
\end{equation}
the function $\eta (z)$ must satisfy the equation
\begin{equation}
\eta (z)-\eta \left( e^{i\pi }z\right) =\theta ;\text{\ \ \ \ }\eta (z)=-%
\dfrac{\theta }{\pi }\arg z=-\dfrac{\theta }{\pi }\arctan \dfrac{y}{x}
\end{equation}
The transformation of the wave function (115) must be accompanied by the
transformation of the electromagnetic field potentials $A_{\mu } $ [44]
\begin{eqnarray}
\dfrac{e}{\hbar c}A_{\mu } &\rightarrow &\dfrac{e}{\hbar c}A_{\mu }+\partial
_{\mu }\eta (z)= \\
&=&\dfrac{e}{\hbar c}\left( A_{\mu }+a_{\mu }\right) ;\text{\ \ }\mu =0,1,2
\notag
\end{eqnarray}
In such a way side by side with the electromagnetic potential $A_{\mu } $
supplementary gauge potential $a_{\mu } $ created by the vortices appears.
\begin{eqnarray}
\dfrac{e}{\hbar c}a_{\mu }(\vec{r}) &=&\partial _{\mu }\eta (\vec{r});
\notag \\
a_{\mu }(\vec{r}) &=&\dfrac{\hbar c}{e}\partial _{\mu }\eta (\vec{r})=-%
\dfrac{\hbar c\theta }{\pi e}\partial _{\mu }\arctan \dfrac{y}{x}
\end{eqnarray}
The statistical gauge vector potential has the expression
\begin{gather}
\vec{a}=\dfrac{\hbar c\theta }{\pi e}Curl\ln r=\dfrac{\hbar c\theta }{\pi e}%
\vec{\triangledown}\times \ln r  \notag \\
a_{i}=\dfrac{\hbar c\theta }{\pi e}\varepsilon ^{ij}\partial _{j}\ln r;\text{%
\ }i,j=1,2
\end{gather}
This vector potential is created by the vortices arising near each particle.
It leads to magnetic field strength [41]
\begin{gather}
b(r)=Curl\vec{a}(r)=\vec{\triangledown}\times \vec{a}(r)=\varepsilon
^{ij}\partial _{i}a_{j}=  \notag \\
=-\dfrac{\hbar c\theta }{e\pi }\Delta \ln r=-\dfrac{2\hbar c\theta }{e}%
\delta ^{(2)}(\vec{r}); \\
\dfrac{\Delta \ln r}{2\pi }=\delta ^{(2)}(\vec{r});  \notag
\end{gather}
The magnetic flux created by this magnetic field is
\begin{equation}
\int b(r)d^{2}\vec{r}=-\dfrac{2\pi \hbar c\theta }{e}=-\dfrac{\theta }{\pi }%
\phi _{0};\phi _{0}=\dfrac{hc}{e}
\end{equation}
It equals to $-(2n+1)\phi _{0} $, when the initial particles described by
the function $\psi (z)$ were fermions, and equals to $-2n\phi _{0} $ for
boson wave function $\psi (z)$. This result shows that the initial fermion
particles each of them attaching an odd number of flux quanta transform
themselves into a composite bosons described by the new wave function $\phi
(z)$ which obeys to Bose statistics. The effective mass m and charge e
remain the same at least in the given approximation but the composition and
statistics of the final quasiparticles are changed. It is said that the
electron attached an odd number of flux tubes, though in reality such tubes
do not exist. We can say that in our case the initial particles are fermions
or electrons whereas the final quasiparticles are bosons. Formula (121) may
be generalized for any number of particles, which create in a common way the
resultant magnetic field
\begin{equation}
b(r)=-\dfrac{2\hbar c\theta }{e}\sum\limits_{i=1}^{N}\delta ^{2}(\vec{r}-%
\vec{r}_{i})=-\dfrac{2\theta \hbar c}{e}\rho (\vec{r})
\end{equation}
where $\rho (\vec{r} )$ is the density of the particles.\newline
As was mentioned above, Zhang, Hanson and Kielson [42] have generalized the
Ginzburg-Landau theory introducing into the Lagrangian a supplementary Chern
- Simoms [35] term related with the influence of the statistical gauge
field. The Lagrangian of the Ginzburg-Landau-Chern-Simons (GLCS) theory in
the form presented by Enger [44] looks as
\begin{eqnarray}
L &=&i\hbar \phi ^{\ast }\left( \partial _{t}+\dfrac{ie}{\hbar c}\left(
A_{0}+a_{0}\right) \right) \phi +  \notag \\
&&+\dfrac{\hbar ^{2}}{2m}\phi ^{\ast }\left( \vec{\triangledown}+\dfrac{ie}{%
\hbar c}\left( \vec{A}+\vec{a}\right) \right) ^{2}\phi - \\
&&-\dfrac{\lambda }{2}\left( \phi ^{\ast }\phi -\rho _{0}\right) ^{2}+\dfrac{%
\mu }{2}e^{\mu \nu \sigma }a_{\mu }\partial _{\nu }a_{\sigma }  \notag
\end{eqnarray}%
\hspace{5mm} Here the following denotations were used: $\mu ,\nu ,\sigma
=0,1,2;$ $\partial _{0}=\partial _{t}$, $\partial _{i}=\left\{ \partial
_{1}=\partial _{x},\text{\ }\partial _{2}=\partial _{y}\right\} $. The
tensor $\varepsilon ^{\mu \nu \sigma }$ has the nonzero components only for
different values of $\mu ,\nu ,\sigma $. They change the signs at any
permutations of two indexes as follows:
\begin{eqnarray}
\varepsilon ^{012} &=&1,\text{\ }\varepsilon ^{021}=-1, \\
\text{\ }\varepsilon ^{\text{102}} &=&-1,\text{\ }\varepsilon ^{120}=1\text{
}etc.  \notag
\end{eqnarray}
The external electromagnetic 2D vector potential $\vec{A} $ and the scalar
potential $A_{0} $ are taken as $A_{\mu } =\left( A_{0} ,\vec{A} \right) $. $%
a_{\mu } $ is the statistical gauge potential with three components. Two of
them $\vec{a} =(a_{1} ,a_{2} )$ generate the statistical ``magnetic'' field
and the third component $a_{0} $ gives rise to the statistical ``electric''
field. Two parameters m and e of the Lagragian (124) are the effective mass
and charge of the final type quasiparticles obeying the Bose statistics.
They can differ from the mass and charge of the initial particles. The $%
\lambda $ and $\rho _{0} $ parameters are typical for the G-L theory, while $%
\mu $ is the Chern-Simons parameter [67]. Variations of (51) with respect to
$\phi ^{*} $ leads to the nonlinear Shrodinger equation:
\begin{eqnarray}
&&\left[ i\hbar \partial _{t}-\dfrac{e}{c}\left( A_{0}+a_{0}\right) \right]
\phi   \notag \\
&=&-\dfrac{\hbar ^{2}}{2m}\left[ \vec{\triangledown}+\dfrac{ie}{\hbar c}%
\left( \vec{A}+\vec{a}\right) \right] ^{2}\phi - \\
&&-\lambda \left( \phi ^{\ast }\phi -\rho _{0}\right) \phi   \notag
\end{eqnarray}
The variation of (49) with respect to $a_{0} $ gives
\begin{equation}
\mu \varepsilon ^{ij}\partial _{i}a_{j}=e\phi ^{\ast }\phi =e\rho
\end{equation}
which can be transcribed as
\begin{equation}
\mu Curl\vec{a}=\mu \vec{\triangledown}\times \vec{a}=\mu b=e\rho
\end{equation}
Parameter $\mu $ can be determined comparing it with the expression (119)
\begin{equation}
\mu =-\dfrac{2\theta \hbar c}{e^{2}}
\end{equation}
For the initial fermion particles with $\theta =\pi (2n+1)$ the parameter $%
\mu $ of the Lagrangian (124) equals to $-\dfrac{2\pi \hbar c}{e^{2} }
(2n+1)=-(2n+1)\dfrac{\phi _{0} }{e} $. The third variation with respect to $%
a_{i} $ gives
\begin{equation}
\mu \varepsilon ^{ij}\left( \vec{\triangledown}a_{0}-\partial _{t}\vec{a}%
\right) =e\vec{j}(r)
\end{equation}
where $\vec{j} (r)$ is the current density 2D vector determined as
\begin{eqnarray}
\vec{j} =\dfrac{\hbar }{2mi}\{\phi ^{\ast }\left( \vec{\triangledown}-%
\dfrac{ie}{\hbar c}\left( \vec{A}+\vec{a}\right) \right) \phi-   \notag \\
-\phi \left( \vec{\triangledown}+\dfrac{ie}{\hbar c}\left( \vec{A}+\vec{a}%
\right) \right) \phi ^{\ast }\}
\end{eqnarray}
The equation (130) states that the statistical "electric" field
\begin{equation}
\vec{\varepsilon}=-\vec{\triangledown}a_{0}+\dfrac{1}{c}\dfrac{\partial }{%
\partial t}\vec{a}
\end{equation}
is related with the particle current density $\vec{j} (r)$.\newline
The energy density of the GLCS system in the purely static external magnetic
field $(A_{0} =0)$ equals to
\begin{equation}
E=\dfrac{\hbar ^{2}}{2m}\left\vert \left( \vec{\triangledown}-\dfrac{ie}{%
\hbar c}\left( \vec{A}+\vec{a}\right) \right) \phi \right\vert ^{2}+\dfrac{%
\lambda }{2}(\rho -\rho _{0})^{2}
\end{equation}
A simple solution to these equations can be obtained by setting
\begin{equation}
\phi (\vec{r})=\sqrt{\rho _{0}}e^{iS(\vec{r})}
\end{equation}
It must obey the equation
\begin{equation}
\vec{\triangledown}S+\dfrac{e}{\hbar c}\left( \vec{A}+\vec{a}\right) =0
\end{equation}
In the case $S=Const$ we have
\begin{equation}
\vec{A}+\vec{a}=0
\end{equation}
It means that the corresponding magnetic fields $B=\vec{\triangledown }
\times \vec{A} $ and $b=\vec{\triangledown } \times \vec{a} $ cancel each
other. The final quasiparticles named as composite bosons [68-70] do not
feel the net magnetic field and behave as bosons interacting with each-other
via the $\phi ^{4} $ type interaction. The notion of composite particles
(CPs) consisting from electrons and attached magnetic flux quanta was
introduced firstly by Wilczek [68].\newline
As was mentioned above, the collective elementary excitations of the
described ground state are the plane waves. Their dispersion law has a gap,
and this means that the system is an incompressible quantum liquid, which
can not be excited by a very small perturbation [69, 70].

Above we have discussed the case when the initial wave function
$\psi (z)$ with Fermi statistics was transformed into another wave function $%
\phi (z)$ obeying Bose statistics using a singular gauge transformation.
Read [40, 71-73] investigated the system of 2D charged bosons interacting
with a transverse magnetic field and between themselves. The filling factor
of the LLL was supposed to be one. It means that there is one flux quantum
for each particle. Following Read [40, 71-73] it is equivalently to say that
there exists one vortex for each particle. In this case the vortex has the
charge of opposite sign in comparison with the charged boson and the Fermi
statistics. Now the gauge transformation attaching one vortex to each
charged boson will create composite particles with resulting charge zero and
with Fermi statistics. The neutral composite fermions (CFs) will move in
zero magnetic field. Such system can be described in the frame of the
Fermi-liquid-theory. Another variant proposed by Halperin, Lee and Read [39]
was considered in Section 5. The starting Hamiltonian describes the
electrons forming a 2D electron gas (2DEG) with filling factor $\nu $ of the
LLL equal to one half $(\nu =1/2)$. Now for each electron there are two flux
quanta or two vortices each of them having the charge $-e/2$ and Fermi
statistics. Two vortices are equivalent to one 2-fold vortex with charge $-e$
and Bose statistics. The gauge transformation of the wave function will
transform the initial charged electrons into the composite neutral fermions
each of them consisting from one electron and 2-fold vortex. The Hamiltonian
of the system will be changed because instead of the external magnetic field
will appear a supplementary gauge magnetic field, which in well definite
conditions cancels exactly the external magnetic field. The initial charged
fermions were converted in neutral CFs moving in a zero resulting magnetic
field. The fictitious Chern-Simons ``magnetic'' field created by the
vortices being averaged in the mean-field approximation cancels exactly the
external magnetic field only in statistical sense and under the definite
conditions. It happens when the mean density of the electrons corresponds to
the fractional integer filling factor. In the present example with $\nu =1/2$
the gauge transformation does not modify the statistics of the composite
particles (CPs). As earlier, they are neutral CFs in a zero magnetic field.
The singular gauge transformations were firstly introduced by Wilczek [68].

The single-particle elementary excitations appear in the form
of the fractionally charged vortices. They are fermions and have finite
creation energy as was underlined by Girvin [52], and Girvin and MacDonald
[53]. Read [39, 40, 71-73] argued that the ground states of the systems in
the condition of FQHE with different fractional integer filling factors $\nu
=1/m$ with $m=1,2,3,...$ contain electrons bound to vortices, since such
binding lowers the system's energy. A m-fold vortex carriers a charge $-e\nu
m$ in the fluid, where $e$ is the electron charge $e=-\left| e\right| $. The
electron-m-vortex composite, named as CP, at $\nu =1/m$ has a net charge
zero and behaves like a particle in a zero magnetic field. The vortex is
sensitive to the density of electrons, which can vary in space and time even
when the external magnetic field and the average filling factor are fixed.
The m-fold vortices are fermions for m odd and bosons for m even. The
composite boson particles can undergo the Bose-Einstein condensation (BEC),
because it minimizes their ``kinetic'' energy. Just the BEC of CBs is the
interpretation of the Laughlin's states [65]. The origin of the ``kinetic''
energy is the potential energy of the interaction between the particles. In
the case of electrons it is the Coulomb electron-electron interaction which
is not canceled by the gauge transformation and CS gauge potential. It is
named as ``kinetic'' because it depends on the wave vector of the operators
(139). The bound objects such as CPs do, in fact, have such an effective
``kinetic'' energy. There is an attraction between an electron and m-fold
vortex. It plays for the electron the role of a correlation quasihole. As
was shown in (139) the CPs may exist in the form of plane waves and the
many-particle wave functions also can be characterized by the wave vector $%
\vec{k} $.

The creation operator in the coordinate representation (56) can
be rewritten in momentum representation as follows
\begin{equation}
\psi ^{\dag }(\vec{k})=\int d^{2}\vec{r}e^{i\vec{k}\vec{r}}\psi ^{\dag }(%
\vec{r})
\end{equation}
A CP with $\vec{k}=0$ would have the electron exactly at the zeroes of the
wave function or in the center of the vortex, whereas the CP with wave
vector $\left\vert k\right\vert \neq 0$ has the electron displaced by the
distance $\left\vert k\right\vert l^{2}$ from their center. One can say that
the electron and its correlation quasihole or in another words the electron
and the m-fold vortex experience a potential $V\left( \left\vert
k\right\vert \right) $ due to the Coulomb interaction of the electron with
other electrons excluded from the vortex core. All these interactions take
place in the presence of the neutralizing background. The electron and the
m-fold vortex experience the magnetic field of the same strength. Both
components of the pair drift in the same direction perpendicular to the
vector connecting their centers, so that their separation remains constant
and equal to $\left\vert k\right\vert l^{2}$. The energy of a pair is $%
V\left( \left\vert k\right\vert \right) $ and its group velocity is $%
\partial V\left( \left\vert k\right\vert \right) /\partial \left\vert
k\right\vert $ [40, 71-73]. We can add that this picture coincides with the
structure of the 2D magnetoexciton, where the energy $V\left( \left\vert
k\right\vert \right) $ equals to the expression $E\left( \left\vert
k\right\vert \right) $ [39]
\begin{eqnarray}
E\left( \left\vert k\right\vert \right)
&=&2\sum\limits_{Q}W_{Q}Sin^{2}\left( \dfrac{\left[ \vec{Q}\times \vec{k}%
\right] _{z}l^{2}}{2}\right) ;  \notag \\
W_{Q} &=&\dfrac{2\pi e^{2}}{\varepsilon _{0}S\left\vert \vec{Q}\right\vert }%
e^{-Q^{2}l^{2}/2}
\end{eqnarray}
Here $\varepsilon _{0} $ is the dielectric constant and S is the layer
surface area.

Girvin, MacDonald and Platzman [54] elaborated the theory of
the collective elementary excitation spectrum in the case of FQHE, which is
closely analogous to the Feynman's theory of superfluid helium. The
predicted spectrum has a gap at $k=0$ and a deep magneto-roton minimum at
finite wavevector, which is a precursor to the gap collapse associated with
Wigner crystal instability. They supposed the existence of only one branch
of the collective elementary excitations spectrum. In this approximation
named as single mode approximation (SMA) they have constructed the wave
functions of the excited states $\phi _{k} $ acting with the operator of the
particle density $\hat{\rho } _{k} $ on the ground state wave function $\psi
_{g} $ in the form $\phi _{k} =\hat{\rho } _{k} \psi _{g} $. They determined
the energy of the excited state $\Delta(k)$ as
\begin{eqnarray}
\Delta (k) =\dfrac{\left\langle \phi _{k}|(H-E_{0})| \phi
_{k}\right\rangle }{\left\langle \phi _{k}|\phi
_{k}\right\rangle }=  \notag \\
=\dfrac{\left\langle \psi _{g}| \rho _{k}^{\dag }\left[ H_{0},\rho
_{k}\right] | \psi _{g}\right\rangle }{\left\langle \psi _{g}|
\rho _{k}^{\dag }\rho _{k}| \psi _{g}\right\rangle }=\dfrac{f(k)}{s(k)}
\end{eqnarray}
where $f(k)$ is the
oscillator strength and $s(k)$ is the static structure factor. The total
oscillator strength sum is saturated by the cyclotron mode contribution, and
$f(k)$ has a dependence of the type $\left\vert k\right\vert ^{4}$. As was
established by Lee and Zhang [74] the influence and the contribution of the
quantum vortices to the dynamical and static structure factors is important.
It leads to dependence $s(k)\sim \left\vert k\right\vert ^{4}$ at $%
k\rightarrow 0$. In this case $\Delta (k)$ has a gap. Neglecting the
influence of the quantum vortices the dependence $s(k)$ is proportional to $%
\left\vert k\right\vert ^{2}$ and the energy spectrum is gapless $\Delta
(k)\approx k^{2}$ at $k\rightarrow 0$ as a Goldstone mode [2]. In
conclusion, for the FQHE in the case $\nu <1$ with fractionally filled
Landau level the Pauli principle no longer excludes the low-lying
intra-Landau-level excitations. They exist side by side with the
inter-Landau-level excitations. The last excitations have a cyclotron energy
gap [75].

We are studying a coplanar electron-hole (e-h) system with
electrons in conduction band and with holes in valence band in a strong
perpendicular magnetic field. Previously such system has been studied in a
series of papers [9, 22-25, 27, 76]. Most of them were dedicated to the
theory of 2D magnetoexcitons. All the same, there were papers dedicate to
another aspects of these systems. For example MacDonald, Rezayi and Keller
[77] as well as Joglekar and MacDonald [78] have discussed the
photoluminescence (PL) spectrum in the FQHE regime. It was mentioned that
the PL spectrum does not exhibit anomalies associated with the FQHE. However
when the electron and hole layers were separated a new peak in the PL
spectrum appears, when the filling factor exceeds a fraction $\nu _{0}$ at
which an incompressible quantum liquid occurs. The new peak is separated
from the main spectral features by the quasiparticle-quasihole gap. We are
interested in the distribution of the flux quanta in the case of e-h system
with equal average numbers of electrons and holes $\bar{N}_{e}=\bar{N}_{h}$
with the filling factor $\nu =\bar{N}_{e}/N$, where $N$ is the total number
of flux quanta , where $S$ is the layer surface area and $2\pi l^{2}$ is the
area of the cyclotron orbit. For the fractional integer filling factor there
are an integer number of flux quanta per each e-h pair. The creation of the
vortices in this case is not studied at present time. But one can expect
that in the case of magnetoexcitons they will be neutral, whereas in the
case of pure electron and pure hole vortices their \textquotedblleft
magnetic\textquotedblright\ gauge fields will compensate each other, so that
the charge, the statistics of the particles and the external magnetic field
will remain the same in the mean-field approximation with equal densities of
electrons and holes. Nevertheless due to quantum fluctuations and the
deviations in space and time of the electron and hole densities from their
average values one can expect the influence of the pure electron and hole
quantum vortices on the physics of magnetoexcitons side by side with the
influence of the neutral quantum vortices formed by the magnetoexcitons
themselves. The last quantum vortices determine the
Berezinskii-Kosterlitz-Thouless phase transition [79, 80].
\section{Quantum Hall excitons in bilayer electron systems}
In this section we give a short review of the Bose-Einstein
condensation (BEC) of the quantum Hall excitons(QHExs) arising in the
bilayer electron systems under the conditions of the quantum Hall effect
(QHE) at one half filling factor $\nu =1/2$ for each layer and the total
filling factor for two layers equal to unity $\nu _{t} =1$. The purpose is
to compare this phenomenon with the case of BEC of two-dimensional (2D)
magnetoexcitons. Such comparison will give a better understanding of the
underlying physics and allows to verify the accuracy of the made
approximations. \newline
\hspace{5mm} In the Ref.[81] Fertig investigated the energy spectrum of a
bilayer electron systems in a strong perpendicular magnetic field and
introduced the concept of the interlayer phase coherence of the electron
states in two adjacent layers, which leads to the model of quantum Hall
excitons under the condition of their BEC. Unexpectedly a strong evidence of
exciton BEC was ultimately found in such surprising system as a double layer
2D electron system at a high magnetic field [82]. In the QHE regime the
excitons consist of electrons in the lowest Landau level (LLL) of the
conduction band of one layer being bound to the holes which appear in the
LLL of the conduction band in another layer. The formation of such unusual
holes is due to the possibility to consider the half-filled LLL by electrons
of the conduction band, for example, of the first layer as being completely
filled by electrons with filling factor $\nu =1$ and simultaneously being
half-filling by holes in the same conduction band. The full-filling
electrons of the first layer are considered as being compensated by the
impurity doped adjacent layer and the theoretical model takes into account
only the holes in the first layer and the electrons in the second layer.
Both components belong to the LLLs of the same conduction band and are
characterized by a half-filling factor for each of them. This new type of
excitons named QHExs appears whenever the temperature and the layer
separation are small enough and the total density $n_{t} $ of electrons in
the double layer system equals to the degeneracy $\dfrac{eB}{hc} =n_{t} =%
\dfrac{1}{2\pi l^{2} } $ where $l$ is the magnetic length. The total filling
factor $\nu _{t} =n_{t} 2\pi l^{2} $ equals unity. The new collective
electronic state introduced by Fertig [81] exhibits several dramatic
electrical transport properties revealed in Ref. [83-85]. As was mentioned
in [82] the BEC of the QHExs reflects the spontaneously broken $U(1)$
symmetry in which the electrons are no-longer confined to one layer or to
the other, but instead they reside in a coherent linear combination of the
two layers. This interlayer phase coherence develops only when the effective
interlayer separation $d/l$ is less than a critical value $(d/l)_{c} $. At
large $d/l$ the bilayer system behaves qualitatively like the independent 2D
electron systems. Following [86] this new state can be distinguished as a
Fermi liquid state of composite fermions. It is unique because unlike other
QH states it possesses a broken symmetry in the absence of the interlayer
tunneling. It can be viewed as a pseudospin ferromagnet with the pseudospin
encoding the layer degrees of freedom or as an exciton BEC with QHExs formed
from electrons and holes confined to different layers.\newline
\hspace{5mm} There are two energy scales in the double-layer systems. One is
the potential energy V between the electrons in different layers. The second
is the energy gap $\Delta _{SAS} $ between the symmetric and asymmetric
states of electrons in two layers measuring the tunneling amplitude between
them. The capability of tuning the strength of the interlayer interaction by
changing the gate voltage provides the opportunity to explore the $\nu _{t}
=1$ system through its transformation between the weak and strong
interaction limits and to study the phase transitions between the
compressible Fermi liquid and the incompressible QH states as a function of $%
d/l$ [86]. In the most theoretical investigations of the QHExs, except the
paper [86], the simplifying assumption of the fully spin polarized electrons
was used. Below, in our discussions the Zeeman energy will be not included.
Following the Ref. [87], in the absence of the interlayer tunneling there
are two $U(1)$ symmetries. One is associated with the conservation of the
total electric charge $N_{1} +N_{2} $, where $N_{1} $ and $N_{2} $ are the
numbers of electrons in two layers, and the other is related with the
conservation law of $N_{1} -N_{2} $. For these conditions the gapless mode
appears. It is the Nambu-Goldstone(NG) mode arising from the broken $U(1)$
symmetry associated with $N_{1} -N_{2} $ and characterized by the
off-diagonal long-range order in the tunneling operator $a_{1p}^{\dag }
a_{2q} $, where $a_{1p} $ and $a_{2q} $ are the electron annihilation
operators in the LLLs of the conduction band of two layers. \newline
\hspace{5mm} Within the mean-field effective theory the appearance of the
gapless mode may be attributed to the coherent fluctuation of the electron
flux and density describing the relative fluctuations of the electron
densities in two layers. At finite interlayer tunneling the number $N_{1}
-N_{2} $ is no-longer conserved. As well, the currents in each layer are
no-longer separately conserved. \newline
\hspace{5mm} The collective excitation spectrum of the two-layer electron
system with $\nu _{t} =1$ was investigated by Fertig [81] on the basis of
the theoretical model without tunneling but with different interlayer
separation including $d=0$ and taking into account that at $d>0$ the Coulomb
interlayer electron-electron interaction is smaller than the intralayer
interaction. The ground state wave function proposed by Fertig [81]
introduces the interlayer phase coherence reflecting a new state, in which
the electrons are no-longer confined to one layer or to another, but instead
they reside in coherent linear combinations of the two layer states as
follows
\begin{eqnarray}
\left\vert \psi \right\rangle  =\prod\limits_{t}\left( ua_{1t}^{\dag
}+va_{2t}^{\dag }\right) \left\vert 0\right\rangle ,  \notag \\
u^{2}+v^{2} =1
\end{eqnarray}
The lowest levels of the Landau quantization in the Landau gauge are characterized by
the quantum number $n=0$ and the uni-dimensional wave number $t$, with $%
\left. 0\right\rangle $ being the vacuum state. The equality $\text{\ }%
u^{2}=v^{2}=1/2$ reflects the half-filling of the LLL in each layer.
Introducing the hole operator $d_{t}^{\dag },d_{t}$ for the first layer
instead of the operators $a_{1t}^{\dag }$ and $a_{1t}$ the function (144)
was transcribed in the form
\begin{gather}
\left\vert \psi \right\rangle =\prod\limits_{t}\left( u+va_{t}^{\dag
}d_{-t}^{\dag }\right) \left\vert \psi _{0}\right\rangle ,\text{\ }  \notag
\\
\left\vert \psi _{0}\right\rangle =\prod\limits_{t}a_{1t}^{\dag }\left\vert
0\right\rangle  \\
a_{2t}=a_{t},\text{\ }a_{2t}^{\dag }=a_{t}^{\dag },\text{\ }%
a_{1t}=d_{-t}^{\dag },\text{\ }a_{1t}^{\dag }=d_{-t}  \notag
\end{gather}
The operators $a_{t}^{\dag }d_{-t}^{\dag }$ create the electron-hole pairs with
total wave vector equal to zero. The wave function (141) can be interpreted
as describing the BEC of the QHExs. This model is similar to the case of BEC
of 2D magnetoexcitons studied in Refs. [22-24, 76]. In the last case the
holes were formed in the frame of the valence band.

The valley-density two-particle integral operators introduced
in Ref. [81] in the electron-hole representation are
\begin{gather}
\rho ^{\pm }(q)=\sum\limits_{t}e^{iq_{y}tl^{2}}[d_{-\dfrac{q_{x}}{2}%
-t}^{\dag }d_{\dfrac{q_{x}}{2}-t}+a_{-\dfrac{q_{x}}{2}+t}^{\dag }a_{\dfrac{%
q_{x}}{2}+t}\pm   \notag \\
\pm (d_{\dfrac{q_{x}}{2}-t}a_{\dfrac{q_{x}}{2}+t}-a_{-\dfrac{q_{x}}{2}%
+t}^{\dag }d_{-\dfrac{q_{x}}{2}-t}^{\dag })]; \\
\rho _{z}(q)=\sum\limits_{t}e^{iq_{y}tl^{2}}(d_{\dfrac{q_{x}}{2}-t}a_{\dfrac{%
q_{x}}{2}+t}+a_{-\dfrac{q_{x}}{2}+t}^{\dag }d_{-\dfrac{q_{x}}{2}-t}^{\dag });
\notag \\
\rho _{F}(q)=\sum\limits_{t}e^{iq_{y}tl^{2}}(a_{-\dfrac{q_{x}}{2}+t}^{\dag
}a_{\dfrac{q_{x}}{2}+t}-d_{-\dfrac{q_{x}}{2}-t}^{\dag }d_{\dfrac{q_{x}}{2}%
-t}).  \notag
\end{gather}
We introduce our designations for the
exciton, optical and acoustical plasmon operators with holes in the
conduction band, as follows
\begin{gather}
\rho (q)=\sum\limits_{t}e^{iq_{y}tl^{2}}[d_{-\dfrac{q_{x}}{2}-t}^{\dag }d_{%
\dfrac{q_{x}}{2}-t}+a_{-\dfrac{q_{x}}{2}+t}^{\dag }a_{\dfrac{q_{x}}{2}+t}]
\notag \\
D(q)=\sum\limits_{t}e^{iq_{y}tl^{2}}[a_{-\dfrac{q_{x}}{2}+t}^{\dag }a_{%
\dfrac{q_{x}}{2}+t}-d_{-\dfrac{q_{x}}{2}-t}^{\dag }d_{\dfrac{q_{x}}{2}-t}]
\notag \\
d(q)=\dfrac{1}{\sqrt{N}}\sum\limits_{t}e^{iq_{y}tl^{2}}d_{\dfrac{q_{x}}{2}%
-t}a_{\dfrac{q_{x}}{2}+t} \\
d^{\dag }(q)=\dfrac{1}{\sqrt{N}}\sum\limits_{t}e^{-iq_{y}tl^{2}}a_{\dfrac{%
q_{x}}{2}+t}^{\dag }d_{\dfrac{q_{x}}{2}-t}^{\dag }  \notag
\end{gather}
In the case of holes in the valence band there are opposite signs in the
expressions for $\rho (q)$ and $D(q)$.
The relations between two sets of operators are
\begin{gather}
\rho ^{\pm }(q)=\rho (q)\pm \sqrt{N}(d(q)-d^{\dag }(-q))  \notag \\
\rho _{z}(q)=\sqrt{N}(d(q)+d^{\dag }(-q)) \\
\rho _{F}(q)=D(q)  \notag
\end{gather}
The response functions were introduced as follows [81]
\begin{gather}
\chi _{\pm }(q,\omega )=-i\int\limits_{0}^{\infty }dte^{i\omega
t}\left\langle \left[ \rho ^{\mp }(q,t),\rho ^{\pm }(-q,0)\right]
\right\rangle   \notag \\
\chi _{z}(q,\omega )=-i\int\limits_{0}^{\infty }dte^{i\omega
t}\left\langle \left[ \rho _{z}(q,t),\rho _{z}(-q,0)\right] \right\rangle  \\
\chi _{F}(q,\omega )=-i\int\limits_{0}^{\infty }dte^{i\omega
t}\left\langle \left[ \rho _{F}(q,t),\rho _{F}(-q,0)\right] \right\rangle
\notag
\end{gather}
The poles of these functions represent the excitations of the system. The
excitations may be thought as a valley-density waves or pseudospin density
waves. The calculation of the response functions were effectuated by Fertig
[81] using the diagrammatic expansion elaborated by Kallin and Halperin
[75]. The response functions were written in terms of vertex functions,
Green's functions and self-energy parts. This approximation is shown by the
diagrams in Fig 3a Ref. [81] neglecting the diagrams which contain
supplementary the e-h bubbles. Their contribution is negligible only when
the excited Landau levels (ELLs) are taken into account and the bubbles have
an energy denominator $\hbar \omega _{c}$, where $\omega _{c}$ is the
cyclotron frequency increasing in the strong field limit. A self-consistent
calculation of the vertex function including the bubbles is quite difficult.
Below we will present the results obtained in Ref. [81] for the energy
spectrum of the collective elementary excitations.

For $d=0$ the interaction Hamiltonian is invariant under the
unitary transformation $SU(2)$. The specific choice of the ground state wave
function (141) is a broken symmetry state and one expects the appearance of
a NG mode. At $d=0$ the NG mode has a dispersion relation $\omega (k)\sim
k^{2} $ for the small $k$. For $d>0$ the problem can be mapped onto an
equivalent spin system with linear dispersion relation at small wave
vectors. The NG mode at $d>0$ has a linear dispersion law with a slope
dependence on $d$, which is similar with that of the acoustical plasmon mode
of a two-layer system in the absence of the magnetic field [88]. To better
understand this result one may recall the BEC interpretation of the ground
state wave function. Indeed. at $d>0$ the inter-layer electron-hole Coulomb
attraction is smaller than the intralayer electron-electron and hole-hole
repulsions, that leads to a resultant repulsion in the system and to the
transformation of the parabolic dispersion law into the linear one at small
values of wave vectors as in the Bogoliubov theory of superfluidity of the
Bose gas [1]. At $kl$ of the order unity the dispersion law of Ref.
[81]develops a dip at certain critical values of $d$, indicating that the
system trends to undergo a phase transition.
Another considerations concerning of the gapless modes in the
FQHE of multicomponent fermions can be found in Ref. [89, 90]. The above
branch of the energy spectrum corresponds to the response function $\chi
_{\pm } (q,\omega )$. The operators $\rho ^{\pm } (q)$ describe two
superpositions of the optical plasmon and exciton mode operators. There are
two other operators $\rho _{z} (q)$ and $\rho _{F} (q)$ which describe the
pure exciton modes and the acoustical plasmon mode. As was established in
Ref. [81] the last two response functions $\chi _{z} (q,\omega )$ and $\chi
_{F} (q,\omega )$ have no poles when the excitations in the frame of the
LLLs are considered. The excitations associated with these functions are
considered to be higher in energy than the NG mode discussed above by an
amount of energy of the order $\hbar \omega _{c} $. It means that the pure
exciton and acoustical plasmon modes in the system of BEC-ed QHExs cannot be
described by the NG gapless modes. As will be shown in the next section, in
similar case of the BEC of coplanar 2D magnetoexcitons the optical plasmon
branch is also the unique NG mode. The exciton branches (energy and
quasienergy) of the spectrum have the gaps in the point $k=0$, the
roton-type behavior at intermediary values of the wave vectors and a
saturation dependences at $k\rightarrow \infty $. At the same time the
acoustical plasmon branch in the case of magnetoexcitons in the range of
small wave vectors reveals the absolute instability. Its values are pure
imaginary. In the case of BEC of magnetoexcitons there is one NG optical
plasmon mode, two gapped exciton modes and one unstable acoustical plasmon
mode. This agrees qualitatively with the results obtained by Fertig [81] in
the case of BEC of QHExs where a single gapless NG mode of the optical
plasmon type was revealed while the other modes of the spectrum were not
identified at infinitesimal energies.
\section{True, quasi and unstable Nambu-Goldstone modes of the Bose-Einstein condensed coplanar magnetoexcitons}
In this section we will present the results following the Refs.
[30, 31] for the energy spectrum of the collective elementary excitations
arising above the ground state of the Bose-Einstein condensed coplanar
magnetoexcitons.

The full Hamiltonian describing the interaction of electrons
and holes lying on the LLLs is:
\begin{equation}
H=H_{Coul}+H_{Suppl}
\end{equation}
Where $%
H_{Coul}$ is the Hamiltonian of the Coulomb interaction of the electrons and
holes lying on their LLLs:
\begin{eqnarray}
\hat{H}_{Coul} =\dfrac{1}{2}\sum\limits_{\vec{Q}}W_{\vec{Q}}\left[ \hat{%
\rho}(\vec{Q})\hat{\rho}(-\vec{Q})-\hat{N}_{e}-\hat{N}_{h}\right] -  \notag
\\
-\mu _{e}\hat{N}_{e}-\mu _{h}\hat{N}_{h}
\end{eqnarray}
and $H_{Suppl} $ is the supplementary indirect interactions between
electrons and holes, which appear due to the simultaneous virtual quantum
transitions of two particles from the LLLs to excited Landau levels (ELLs)
and their return back during the Coulomb scattering processes. The
expression for this interaction was obtained in Ref. [25] and has the form:
\begin{eqnarray}
H_{\text{suppl}} &=&\dfrac{1}{2}B_{i-i}\widehat{N}-\dfrac{1}{4N}%
\sum\limits_{Q}V(Q)\hat{\rho}(\vec{Q})\hat{\rho}(-\vec{Q})-  \notag \\
&&-\dfrac{1}{4N}\sum\limits_{Q}U(Q)\hat{D}(\vec{Q})\hat{D}(-\vec{Q})
\end{eqnarray}
Here $\hat{\rho}(\vec{Q})$ are the
density fluctuation operators expressed through the electron $\hat{\rho}_{e}(%
\vec{Q})$ and hole $\hat{\rho}_{h}(\vec{Q})$ density operators as follows:%
\begin{gather}
\widehat{\rho }_{e}(\overrightarrow{Q})=\sum\limits_{t}e^{iQ_{y}tl^{2}}a_{t-%
\dfrac{Q_{x}}{2}}^{\dag }a_{t+\dfrac{Q_{x}}{2}};  \notag \\
\widehat{\rho }_{h}(\overrightarrow{Q})=\sum\limits_{t}e^{iQ_{y}tl^{2}}b_{t+%
\dfrac{Q_{x}}{2}}^{\dag }b_{t-\dfrac{Q_{x}}{2}}  \notag \\
\hat{\rho}(\vec{Q})=\hat{\rho}_{e}(\vec{Q})-\hat{\rho}_{h}(-\vec{Q}); \\
\hat{D}(\vec{Q})=\hat{\rho}_{e}(\vec{Q})+\hat{\rho}_{h}(-\vec{Q});  \notag \\
\hat{N}_{e}=\widehat{\rho }_{e}(0);\text{\ \ }\hat{N}_{h}=\widehat{\rho }%
_{h}(0);\text{\ \ }  \notag \\
\hat{N}=\hat{N}_{e}+\hat{N}_{h};\text{\ }W_{\vec{Q}}=\dfrac{2\pi e^{2}}{%
\varepsilon _{0}S\left\vert \vec{Q}\right\vert }e^{-Q^{2}l^{2}/2}  \notag
\end{gather}
The density operators are integral
two-particle operators. They are expressed through the single-particle
creation and annihilation operators $a_{p}^{\dag },a_{p}$ for electrons and $%
b_{p}^{\dag },b_{p}$ for holes. Here, $\varepsilon _{0}$ is the dielectric
constant of the background; $\mu _{e}$ and $\mu _{h}$ are chemical
potentials for electrons and holes, and coefficients $V(Q)$, $U(Q)$ and $%
B_{i-i}$ were calculated in [25, 29].

The starting Hamiltonian (146) has two continuous symmetries.
One is the gauge global symmetry $U(1)$ and another one is the rotational
symmetry $SO(2)$, so that the total symmetry is $U(1)\times SO(2)$. The
gauge symmetry is generated by the operator $\hat{N} $ of the full particle
number, when it commutes with the Hamiltonian. It means that the Hamiltonian
is invariant under the unitary transformation $\hat{U} (\varphi )$ as follows
\begin{eqnarray}
\hat{U}(\varphi )\hat{H}\hat{U}^{-1}(\varphi ) &=&\hat{H}; \\
\hat{U}(\varphi ) &=&e^{i\hat{N}\varphi };\text{\ \ }[\hat{H},\hat{N}]=0
\notag
\end{eqnarray}
The operator $\hat{N} $ is referred to as the symmetry generator. The
rotational symmetry $SO(2)$ is generated by the rotation operator $\hat{C}
_{z} (\varphi )$ which rotates the in-plane wave vectors $\vec{Q} $ on the
arbitrary angle $\varphi $ around $z$ axis, which is perpendicular to the
layer plane and is parallel to the external magnetic field. Coefficients $W_{%
\vec{Q} } $, $U(\vec{Q} )$ and $V(\vec{Q} )$ in formulas (6) and (9) of Ref.
[30] depend on the square wave vector $\vec{Q} $ which is invariant under
the rotations $\hat{C} _{z} (\varphi )$. This fact determines the symmetry $%
SO(2)$ of the Hamiltonian (146). The gauge symmetry of Hamiltonian (146)
after the phase transition to the Bose-Einstein condensation (BEC) state is
broken as it follows from expression (16) of Ref. [30]. In terms of the
Bogoliubov theory of quasiaverages, it contains a supplementary term
proportional to $\tilde{\eta } $. The gauge symmetry is broken because this
term does not commute with operator $\hat{N} $. Moreover, this term is not
invariant under the rotations $\hat{C} _{z} (\varphi )$, because the
in-plane wave vector $\vec{k} $ of the BEC is transformed into another wave
vector rotated by the angle $\varphi $ in comparison with the initial
position. The second continuous symmetry is also broken. Thus, the
installation of the Bose-Einstein condensation state with arbitrary in-plane
wave vector $\vec{k} $ leads to the spontaneous breaking of both continuous
symmetries.

We will consider a more general case of $\vec{k} \neq 0$ taking
the case $\vec{k} =0$ as a limit $\vec{k} \rightarrow 0$ of the cases with
small values $kl$\texttt{<}\texttt{<}1. One should keep in mind that the
supplementary terms in Hamiltonian (146) describing influence of the ELLs
are actual in the range of small value $kl<0.5$. Above we established that
the number of the broken generators (BGs) denoted as $N_{BG} $ equals to two
($N_{BG} =2$).

As discussed in previous papers [22-25, 27, 76, 91, 92], the
breaking of the gauge symmetry of the Hamiltonian (146) can be achieved
using the Keldysh-Kozlov-Kopaev [93] method with the unitary transformation
\begin{equation}
\hat{D}(\sqrt{N_{ex}})=\exp [\sqrt{N_{ex}}(d^{\dag }(\vec{k})-d(\vec{k}))],
\end{equation}
where $%
d^{\dag }(\vec{k})$ and $d(\vec{k})$ are the creation and annihilation
operators of the magnetoexcitons. In the electron-hole representation they
are [22-25, 27, 76, 91, 92]:
\begin{gather}
d^{\dag }(\vec{P})=\dfrac{1}{\sqrt{N}}\sum\limits_{t}e^{-iP_{y}tl^{2}}a_{t+%
\dfrac{P_{x}}{2}}^{\dag }b_{-t+\dfrac{P_{x}}{2}}^{\dag };  \notag \\
d(\vec{P})=\dfrac{1}{\sqrt{N}}\sum\limits_{t}e^{iP_{y}tl^{2}}b_{-t+\dfrac{%
P_{x}}{2}}a_{t+\dfrac{P_{x}}{2}};
\end{gather}
BEC of the
magnetoexcitons leads to the formation of a coherent macroscopic state as a
ground state of the system with wave function
\begin{equation}
\left\vert \psi _{g}(\bar{k})\right\rangle =\hat{D}(\sqrt{N_{ex}})\left\vert
0\right\rangle ;a_{p}\left\vert 0\right\rangle =b_{p}\left\vert
0\right\rangle =0
\end{equation}
Here $\left\vert
0\right\rangle $ is the vacuum state for electrons and holes. In spite of
the fact that we kept arbitrary value of $\vec{k}$, nevertheless our main
goal is the BEC with $\vec{k}=0$ and we will consider the interval $%
0.5>kl\geq 0$. The function (153) will be used to calculate the averages
values of the type $\left\langle D(\vec{Q})D(-\vec{Q})\right\rangle $. The
transformed Hamiltonian (146) looks like:
\begin{equation}
\hat{\mathcal{H}}=D\left( \sqrt{N_{ex}}\right) HD^{\dag }\left( \sqrt{N_{ex}}%
\right)
\end{equation}
and is succeeded, as usual, by the Bogoliubov u-v transformations of the
single-particle Fermi operators
\begin{gather}
\alpha _{p}=\hat{D}\left( \sqrt{N_{ex}}\right) a_{p}\hat{D}^{\dag }\left(
\sqrt{N_{ex}}\right) =  \notag \\
=ua_{p}-v(p-\dfrac{k_{x}}{2})b_{k_{x}-p}^{\dag };  \notag \\
\alpha _{p}\left\vert \psi _{g}(\bar{k})\right\rangle =0; \\
\beta _{p}=\hat{D}\left( \sqrt{N_{ex}}\right) b_{p}\hat{D}^{\dag }\left(
\sqrt{N_{ex}}\right) =  \notag \\
=ub_{p}+v(\dfrac{k_{x}}{2}-p)a_{k_{x}-p}^{\dag };  \notag \\
\beta _{p}\left\vert \psi _{g}(\bar{k})\right\rangle =0;  \notag
\end{gather}
Instead of this traditional way of transforming
the expressions of the starting Hamiltonian (146) and of the integral
two-particle operators (149) and (152), we will use the method proposed by
Bogoliubov in his theory of quasiaverages [1, 51], remaining in the
framework of the original operators. The new variant is completely
equivalent to the previous one, and both of them can be used in different
stages of the calculations. For example, the average values can be
calculated using the wave function (153) and u-v transformations (155),
whereas the equations of motion for the integral two-particle operators can
be simply written in the starting representation.

The Hamiltonian (146) with the broken gauge symmetry in the
lowest approximation has the form
\begin{gather}
\hat{\mathcal{H}}=\dfrac{1}{2}\sum\limits_{\vec{Q}}W_{\vec{Q}}\left[ \rho (%
\vec{Q})\rho (-\vec{Q})-\hat{N}_{e}-\hat{N}_{h}\right] -  \notag \\
-\mu _{e}\hat{N}_{e}-\mu _{h}\hat{N}_{h}+\dfrac{1}{2}B_{i-i}\widehat{N}-
\notag \\
-\dfrac{1}{4N}\sum\limits_{Q}V(Q)\hat{\rho}(\vec{Q})\hat{\rho}(-\vec{Q})- \\
-\dfrac{1}{4N}\sum\limits_{Q}U(Q)\hat{D}(\vec{Q})\hat{D}(-\vec{Q})-  \notag
\\
-\tilde{\eta}\sqrt{N}\left( d^{\dag }(k)+d(k)\right)   \notag
\end{gather}
For simplicity another smaller term of this type proportional to $\tilde{\eta%
}$ was dropped. Here parameter $\tilde{\eta}$, which determines the breaking
of the gauge symmetry, depends on the chemical potential $\mu $ and on the
square root of the density, similar to the case of weakly non-ideal Bose-gas
considered by Bogoliubov [1, 51]. In our case the density is proportional to
the filling factor $\nu =\text{v}^{2}$ and we have:
\begin{eqnarray}
\mu  =\mu _{e}+\mu _{h};\bar{\mu}=\mu +I_{l};N_{ex}=v^{2}N;  \notag \\
\tilde{E}_{ex}(k) =-I_{l}-\Delta (k)+E(k); \\
\tilde{\eta} =(\tilde{E}_{ex}(k)-\mu )v=(E(k)-\Delta (k)-\bar{\mu})v;
\notag \\
E(k) =2\sum\limits_{Q}W_{Q}Sin^{2}\left( \dfrac{[K\times Q]_{z}l^{2}}{2}%
\right) ;  \notag
\end{eqnarray}
The equations of motion for the integral
two-particle operators with wave vectors $\vec{P}\neq 0$ in the special case
of BEC of magnetoexcitons with $\vec{k}=0$ are
\begin{gather}
i\hbar \dfrac{d}{dt}d(\vec{P})=[d(\vec{P}),\hat{\mathcal{H}}];  \notag \\
i\hbar \dfrac{d}{dt}d^{\dag }(-\vec{P})=[d^{\dag }(-\vec{P}),\hat{\mathcal{H}}]
\end{gather}
\begin{gather*}
i\hbar \dfrac{d}{dt}\hat{\rho}(\vec{P})=[\hat{\rho}(\vec{P}),\hat{\mathcal{H}%
}]; \\
i\hbar \dfrac{d}{dt}\hat{D}(\vec{P})=[\hat{D}(\vec{P}),\hat{\mathcal{H}}];
\end{gather*}
Following the equations of motion (158) we introduce four interconnected
retarded Green's functions at $T=0$ [94, 95]
\begin{gather}
G_{11}(\vec{P},t)=\left\langle \left\langle d(\vec{P},t);\hat{X}^{\dag }(%
\vec{P},0)\right\rangle \right\rangle ;  \notag \\
G_{12}(\vec{P},t)=\left\langle \left\langle d^{\dag }(-\vec{P},t);\hat{X}%
^{\dag }(\vec{P},0)\right\rangle \right\rangle ;  \notag \\
G_{13}(\vec{P},t)=\left\langle \left\langle \dfrac{\hat{\rho}(\vec{P},t)}{%
\sqrt{N}};\hat{X}^{\dag }(\vec{P},0)\right\rangle \right\rangle ; \\
G_{14}(\vec{P},t)=\left\langle \left\langle \dfrac{\hat{D}(\vec{P},t)}{\sqrt{%
N}};\hat{X}^{\dag }(\vec{P},0)\right\rangle \right\rangle ;  \notag
\end{gather}
We also need their Fourier transforms $G_{ij} (\vec{P} ,\omega )$, for which
the equations of motion of the type similar to the equations of motion (158)
were obtained. These Green's functions can be named as one operator Green's
functions, because they contain only one two-particle operator of the type $%
d^{\dag } $, $d$, $\rho $, $D$. At the same time, in the right hand side of
the corresponding equations of motion there is a second generation of
two-operator Green's function containing the different products of the
two-particle operators mentioned above. For these operators was derived the
second generation of the equations of motion containing in their right sides
the Green's function of the third generation. They are the three-operator
Green's functions for which it is necessary to derive the third generation
of equations of motion. However, we have to terminate here the evolution of
the infinite chains of equations of motion for multi-operators Green's
function following the procedure proposed by Zubarev [95]. The truncation of
the chains of the equations of motion and the decoupling of the one-operator
Green's functions from the multi-operator Green's functions was achieved
substituting the three operator Green's functions by the one-operator
Green's functions multiplied by the average value of remaining two
operators. The average values were calculated using the ground state wave
function (153) and u-v transformations (155). The Zubarev procedure is
equivalent to a perturbation theory with a small parameter of the type $%
v^{2}(1-v^{2})$, which represent the product of a
filling factor $\nu =v^{2} $ and the phase-space filling
factor $(1-v^{2} )$ reflecting the Pauli exclusion principle.
The closed system of Dyson equations has the form
\begin{eqnarray}
\sum\limits_{j=1}^{4}G_{1j}(\vec{P},\omega )\Sigma _{jk}(\vec{P},\omega )
&=&C_{1k}; \\
k &=&1,2,3,4  \notag
\end{eqnarray}
There are 16 different components of
the self-energy parts $\Sigma _{jk}(\vec{P},\omega )$ forming a $4\times 4$
matrix. Due to the structure of the self-energy parts the cumbersome
dispersion equation can be expressed in general form by the determinant
equation
\begin{equation}
\det \left\vert \Sigma _{ij}(\vec{P},\omega )\right\vert =0
\end{equation}
It splits into two independent equations. One of them concerns only the
optical plasmon branch and has a simple form
\begin{equation}
\Sigma _{33}(\vec{P},\omega )=0
\end{equation}
It does not include the chemical potential $\bar{\mu}$ and the
quasiaverage constant $\tilde{\eta}$. The second equation contains the
self-energy parts $\Sigma _{11}$, $\Sigma _{22}$, $\Sigma _{44}$, $\Sigma
_{14}$, $\Sigma _{41}$, $\Sigma _{24}$ and $\Sigma _{42}$, which include
both parameters $\bar{\mu}$ and $\tilde{\eta}$. The second equation reads
\begin{gather}
\Sigma _{11}(\vec{P};\omega )\Sigma _{22}(\vec{P};\omega )\Sigma _{44}(\vec{P%
};\omega )-  \notag \\
-\Sigma _{41}(\vec{P};\omega )\Sigma _{22}(\vec{P};\omega )\Sigma _{14}(\vec{%
P};\omega )- \\
-\Sigma _{42}(\vec{P};\omega )\Sigma _{11}(\vec{P};\omega )\Sigma _{24}(\vec{%
P};\omega )=0  \notag
\end{gather}
The solution of the equation (162) is
\begin{eqnarray}
&&\left( \hbar \omega (P)\right) ^{2}  \notag \\
&=&\dfrac{\left\langle D(P)D(-P)\right\rangle }{N^{2}}\sum%
\limits_{Q}U(Q)(U(-Q)- \\
&&-U(Q-P))Sin^{2}\left( \dfrac{[P\times Q]_{z}l^{2}}{2}\right)   \notag
\end{eqnarray}
The right hand side of this expression at small
values of $P$ has the dependence $\left\vert P\right\vert ^{4}$ and tends to
saturation at large values of $P$. The optical plasmon branch $\hbar \omega
_{OP}(P)$ has a quadratic dispersion law in the long wavelength limit and
saturates in the range of short wavelengths. It depends on concentration as $%
\sqrt{v^{2}(1-v^{2})}$ what coincides with the concentration
dependencies of 3D plasma $\omega _{p}^{2}=\dfrac{4\pi e^{2}n_{e}}{%
\varepsilon _{0}m}$ [96] and 2D plasma $\omega _{p}^{2}(q)=\dfrac{2\pi
e^{2}n_{s}q}{\varepsilon _{0}m}$ [88], where $n_{e}$ and $n_{s}$ are the
corresponding density of electrons. The supplementary factor $(1-v%
^{2})$ in our case reflects the Pauli exclusion principle and the vanishing
of the plasma oscillations at $\nu =v^{2}=1$. The obtained dispersion
law is shown in Fig.2. Similar dispersion law was obtained for the case of
2D electron-hole liquid (EHL) in a strong perpendicular magnetic field [97],
when the influence of the quantum vortices created by electron and hole
subsystems is compensated exactly. However, the saturation dependencies in
these two cases are completely different. In the case of Bose-Einstein
condensed magnetoexcitons it is determined by the ELLs, whereas in the case
of EHL [97] it is determined by the Coulomb interaction in the frame of the
LLLs.

The acoustical plasmon branch has the dispersion law, which is
completely different from the optical plasmon oscillations. It has an
absolute instability beginning with the small values of wave vector going on
up to the considerable value $pl\approx 2$. In this range of wave vectors,
the optical plasmons have energies which do not exceed the activation energy
$U(0)$. It means that the optical plasmons containing the opposite-phase
oscillations of the electron and hole subsystems without displacement as a
whole of their center of mass are allowed in the context of the attractive
bath. On the other hand, the in-phase oscillations of the electron and hole
subsystems in the composition of the acoustical plasmons are related to the
displacements of their center of mass. Such displacements can take place
only if their energy exceeds the activation energy $U(P)$. As a result, the
acoustical plasmon branch has an imaginary part represented by the dashed
line and is completely unstable in the region of wave vectors $pl\leq 2$. At
greater values $pl>2$ the energy spectrum is real and nonzero, approaching
to the energy spectrum of the optical plasmons.

In case of 2D magnetoexcitons in the BEC state with small wave
vector $kl<0.5$ described by the Hamiltonian (156), one should take into
account that both continuous symmetries usual for the initial form (146) are
lost. It happened due to the presence of the term $\tilde{\eta}(d_{\vec{k}%
}^{\dag }+d_{\vec{k}}^{{}})$ in the frame of the Bogoliubov theory of the
quasiaverages. Nevertheless the energy of the ground state as well as the
self-energy parts $\Sigma _{ij}(P,\omega )$ were calculated only in the
simplest case of the condensate wave vector $\vec{k}=0$. These expressions
can be relevant also for infinitesimal values of the modulus $\left\vert
\vec{k}\right\vert $ but with a well defined direction. In this case the
symmetry of the ground state will be higher than that of the Hamiltonian
(156), what coincides with the situation described by Georgi and Pais [33].
It is one possible explanation of the quasi-NG modes appearance in the case
of exciton branches of the spectrum. Another possible mechanism of the
gapped modes appearance is the existence of the local gauge symmetry, the
breaking of which leads to the Higgs effect [7]. The interaction of the
electrons with the attached vortices gives rise to a gapped energy spectrum
of the collective elementary excitations as was established in Ref.[42, 54].
The number of the NG modes in the system with many broken continuous
symmetries was determined by the Nielsen and Chadha [17] theorem. It states
that the number of the first-type NG modes $N_{I}$ being accounted once and
the number of the second type NG modes $N_{II}$ being accounted twice equals
or prevails the number of broken generators $N_{BG}$. It looks as follows $%
N_{I}+2N_{II}\geq N_{BG}$. In our case the optical plasmon branch has the
properties of the second-type NG modes. We have $N_{I}=0;$ $N_{II}=1$ and $%
N_{BG}=2$. It leads to the equality $2N_{II}=N_{BG}$. The three branches of
the energy spectrum are represented together in the Fig.2. One of then is a
second-type Nambu-Goldstone(NG) mode describing the optical plasmon-type
excitations, the second branch is the first-type NG mode with absolute
instability describing the acoustical-type excitations and the third branch
is the quasi-NG mode describing the exciton-type collective elementary
excitations of the system.\newline

We can repeat that results obtained in the magnetoexciton
system are similar to those obtained for the system of BEC of the quantum
Hall excitons (QHExs) [81]. In these both models there is only one gapless
Nambu-Goldstone mode between four branches of the energy spectrum. In our
model it is related with the optical plasmon branch, whereas in the case of
QHExs this mode is represented by the superposition of the operators
describing the optical plasmon and exciton modes. In both models the exciton
branches of the spectrum are not gapless and differ from the NG modes. In
our case the exciton energy and quasienergy branches corresponding to normal
and abnormal Green's functions have a gaps in the point $k=0$, a roton-type
segments in the region of intermediary wave vectors $kl\sim 1$ and
saturation-type behaviors at great values of $kl$. In the case of Ref[81]
the excitontype response function $\chi _{z}(q,\omega )$ and the
acoustical-type response function $\chi _{F}(q,\omega )$ have no poles in
the region of small energies in the frame of the LLLs. It was concluded that
the energies of these excitations may be situated at greater values. In our
case the acoustical plasmon branch reveals an absolute instability in the
range of small and intermediary values of k. It means that a real values of
the pole does not exist in the range of small energies, which is similar
with the results of Ref[81]. One can conclude that the qualitative
properties of the energy spectra in both models are similar in spite of the
mentioned differences. It is an additional argument supporting the accuracy
of our calculations, which satisfy to the Nielsen and Chadha theorem [17].

\begin{figure}
\resizebox{0.48\textwidth}{!}{%
  \includegraphics{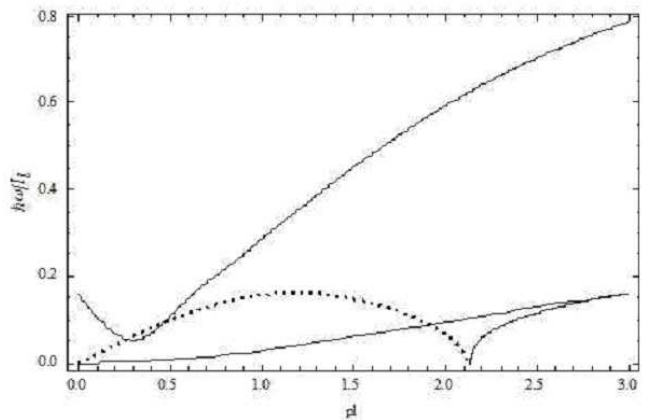}
}
\caption{Three branches of the collective elementary excitations: the exciton-type quasi-NG mode with a gap in the point $pl=0$; the second-type NG mode describing the optical plasmons and the
first-type NG mode with absolute instability (dotted line) describing the
acoustical plasmons.}
\label{fig:2}       
\end{figure}

$pl=0$; the second-type NG mode describing the optical plasmons and the
first-type NG mode with absolute instability (dotted line) describing the
acoustical plasmons.

The result concerning the BEC at $T=0$ are estimates which
describe the real situation at finite temperatures lower than the critical
temperature of the Berezinskii-Kosterlitz-Thouless (BKT) topological phase
transition [79, 80] related with the existence of the vortices and their
clusters such as bound vortex-antivortex pairs. Just the unbinding of these
pairs determines the critical temperature $T_{BKT} =\dfrac{\pi n\hbar ^{2} }{%
2mk_{B} } $, where n is the surface density of the Bose particles and m is
their mass. On one side of the phase transition there is a quasi-ordered
fluid and on the other is a disordered unbounded vortex plasma. Although the
formation of an isolated vortex will not occur at low temperature due its
extensive creation energy, there always can be production of a pair of
vortices with equal and opposite charges since the perturbation produced by
such a pair falls off sufficiently rapidly at large distances so that their
energy is finite [80]. Such topological formations can be easily created by
the thermal fluctuations.

The presence of the vortex clusters makes the previously
infinite homogeneous 2D e-h system to become nonhomogeneous as a whole.
However, the local homogeneity with finite local surface areas can exist
leading to the BEC with finite critical temperature $T_{c} = \frac{2n\pi\hbar^{2}}{mk_{B} \lg(nS)}$ [98]. Instead of an off diagonal
long-range order as in the case of 3D Bose gas in the 2D systems there is
only a long rang correlations, which decays algebraically with distance. In
such a way the quantum vortices promote the BEC and the formation of the
superfluid component of the 2D Bose-gas at finite temperatures and at the
same time the superfluid component is necessary for the formation of the
quantum vortices. It is a self-organization-type situation. The BKT phase
transition is a widely studied phenomenon [99-101].

Attempts to discover experimentally the spontaneous symmetry
breaking in the exciton range of the spectrum and the efforts to evidence
the spontaneous coherence in the 2D excitonic systems are will be considered
in the next section on the basis of the references [102-146].
\section{Spontaneous coherence in 2D excitonic systems}
As was mentioned by Snoke in Refs. [102, 103] recent
experimental efforts of several groups have demonstrated the spontaneous
coherence in polariton systems, which can be viewed as a type of
nonequilibrium BEC. The system of polaritons in the quantum wells embedded
into the microcavity reveals the phenomenon of BEC and superfluity. The
achievements in this field are presented in Ref. [104-110]. In these systems
the polariton lifetime is longer than, but not much longer than the
polariton-polariton scattering time, which leads to the thermalization. By
contrast over past twenty years several groups of investigators represented
by Snoke [111-121], Butov [122-130], Timofeev [131-138], Krivolapchuk
[139-143], Fukuzawa [144-146] and their coworkers have pursued experiments
in double quantum well (DQW) excitonic systems with very long lifetime. In
these systems the indirect excitons (IXs) formed from spatially separated
electrons and holes have dipole moments oriented perpendicularly to the
layers. They are named dipole excitons and their interaction is not a
short-range contact interaction but instead a long-range dipole-dipole
repulsion. We briefly recall the results obtained in Refs. [142, 143].

When analyzing the possibility of BEC in a 2D system it should
be noted that at $T\neq 0$ condensation of a homogeneous 2D gas is
impossible because of destruction of the condensate by thermal fluctuations
[147]. In a 2D system $\rho (E)$ is constant and the integral $%
N=\sum\limits_{k}N_{k}=\underset{0}{\overset{\infty }{\int }}\frac{\rho
(E)dE}{e^{(E-\mu )/k_{B}T}-1}$ would diverge at $\mu \rightarrow 0$ and $%
T\neq 0$ because of the zero denominator at the lower integrating limit and
therefore BEC is impossible here. Physically this fact means that the
maximal occupation of free states $(E>0)$ is infinite. However, if in a 2D
boson system, together with free excitons, there are present some discrete
states (localized states whose existence is caused by the appearance of
fluctuations of the heterointerface potential [148]) $\varepsilon _{0}$, $%
\varepsilon _{1}$ etc, such that $\varepsilon _{0}<\varepsilon _{1}<E=0$,
the situation changes essentially. In this case under increasing number of
bosons in the system the value of chemical potential cannot be arbitrarily
close to the $E=0$ value because of the $N(\varepsilon _{0})\geq 0$
requirement, so $(-\mu )_{\min }=|\varepsilon _{0}|$ and, consequently, the
integral $N$ has a finite value [149]:
\begin{equation}
n_{c}(T)=-\dfrac{mk_{B}T}{2\pi \hbar }\ln (1-\exp ^{-\dfrac{|\varepsilon
_{0}|}{k_{B}T}})
\end{equation}
Therefore, at the moment when n exceeds $n_{c}(T)$, localized states are occupied by a
macroscopic number of particles: $n-n_{c}(T)=n(\varepsilon
_{0})+n(\varepsilon _{1})$. This means that BEC into localized states occurs
in a limited space region. In this sense BEC in a system of 2D bosons,
having a discrete spectrum of energy together with a continuous one,
resembles the experimentally discovered phenomenon of BEC in atoms of alkali
metals in space-limited traps produced by a magnetic field [150].

The obvious advantage of an IX in a DQW as a perspective system
to reveal BEC is the possibility of controlling effectively its radiative
lifetime $\tau _{R} $ with the help of external effects. So, for example, an
electric field Vdc applied to a DQW in the direction of the growth axis of
the structure causes an essential decrease of the overlap of wavefunctions
of an electron and a hole in an IX in the z direction and, as a result, $%
\tau _{R} $ increases significantly (by up to three orders of magnitude
[145]). This allows a more effective cooling of the system to the bath
temperature and, of equal importance, gives an opportunity to increase the
concentration of the IX gas without increasing the pumping density. The
latter circumstance plays an important role in the experiment since it
allows us to decrease heating of the sample by phonons that are inevitably
radiated at relaxation of photoexcited carriers and excitons. Just this
heating of the sample under investigation is often the main cause of the
impossibility of reaching the critical temperature of the boson gas in
experiments that use large optical pumping densities to create the critical
density of bosons having very short lifetimes.

Convincing evidence of the BEC effect would be an appropriate
exciton distribution function over energies (momenta) obtained in an
experiment. In general the exciton distribution function can be determined
in experiment by the form of the phonon replica line in luminescence
spectra, but in our study no phonon replica was observed. This was why in
this paper the nonphonon luminescence line of space IXs was studied.
However, since the intensity of exciton radiation is proportional to the
occupation of radiative states by particles, it indirectly reflects the
distribution function of excitons over the free and localized states, which
both contribute to the formation of the inhomogeneously broadened IX line.
Due to this fact one can hope that studies of IX luminescence will reveal
the BEC effect predicted [149] for a system of 2D bosons which are
distributed over the free and localized states.

A giant (threefold) increase of luminescence intensity of a
part of the spectral profile of the IX line in DQWs of GaAs/Al$_{0}$$_{.}$$%
_{33}$Ga$_{0}$$_{.}$$_{67}$As on changing the temperature of the sample and
the value of the external electric field applied to DQWs was discovered.
Besides that, the luminescence intensity of this part of the spectral
profile of the IX line fluctuated with the characteristic time of tens of
seconds. Such an unusual behaviour of the IX line was regarded as possible
evidence for BEC in a system of 2D bosons placed in fluctuations of
potential formed by heterointerfaces of the sample.\newline
\hspace{5mm} Figure 1 of the Ref. [142] is reproduced in Fig.3. It shows the
luminescence spectra of the DQW, dependent on $V_{dc} $ at $T=1.8K$ and the
density of optical excitation $P=5Wcm^{-2} $. Here, at $V_{dc} =0$ (figure
(a)) the radiation spectrum was close to that of the flat-band case and it
consisted of two lines, DXW and DXN, corresponding to luminescence of direct
excitons (DXs) from the wide and the narrow wells respectively. At nonzero
Vdc (figures (b)--(e)) an indirect regime was achieved (see the inset) when
the IX line took the lowest energetic position in the PL spectra. When Vdc
increased the IX line moved monotonically towards lower energies.

It was noted that in some interval of $V_{dc} $ a giant (up to
threefold) increase (shot) of intensity of a part of the IX spectral line
profile (figure (d)) Ref. [142] occurred. A very important circumstance was
that the intensity shot was absent from the whole investigated interval of
spectral positions of the IX line at any temperature $4.2K\leq T\leq 30K$
and optical pumping densities of $P\leq 5Wcm^{-2} $. Thus the spectral
profile of the IX line corresponding to the case of figure (d) was shown on
a large scale in the figure 3 of Ref. [142] and it is reproduced here in
Fig.4. It has some interesting peculiarities consisting of a narrow intense
line C and of `wings' W having significantly smaller intensity. Measurements
of the temporal evolution of intensities of the C and W components have
shown that the intensity of C, in contrast to that of W, fluctuated in time
(changing threefold) on a characteristic scale of tens of seconds.

The shape of the IX luminescence line in a DQW was
inhomogeneous and was determined by exciton radiation from different space
regions of the DQWs plane, which differed from one another by the thickness
of the QW layers, by fluctuations of barrier composition and by the value of
the local electric field of impurities. The emission intensity of each
spectral fragment of the IX line was proportional to the exciton occupation
of the corresponding space region in the plane of the QW. However, a
spectral region ( $V_{dc} $) where the situation changes essentially at $%
T=1.8K$ and $P=5Wcm^{-2} $ appeared. In this region, shown in figure 3 (d) a
significant (threefold) increase of luminescence intensity of a part of the
spectral profile of the IX line and a consequent increase (1.5-fold) of
integral intensity $I_{IX} $ had been observed. Such a behavior was
anomalous in comparison to the monotonic decrease of $I_{IX} $ (and, most
important, to the absence of the intensity shot) with increasing $V_{dc} $
under other experimental conditions (at $T=1.8K$ and $P<1Wcm^{-2} $; $%
4.2K\leq T\leq 30K$ and any $P\leq 5Wcm^{-2} $ as well). This anomalous
behavior of the IX line indicated that in the case of figure 1 (d) of Ref.
[142] there are much more particles participating in radiation (occupying
states which can radiate) than in the cases of figures 1 (b), (c), (e) of
Ref. [142].

To explain this the BEC model developed for 2D systems [149]
was proposed. It was shown that if in a system of bosons (excitons) there is
a localized state $\varepsilon _{0} $ below the bottom of the exciton band,
the chemical potential of excitons is trapped by the localized level $%
\varepsilon _{0} $ and, as a consequence, the number $n_{c} $ of particles
appears to be finite. Therefore just when the concentration n of excitons in
the system under consideration exceeds $n_{c} $, a macroscopic number $%
n-n_{c} $ of particles comes to the lowest energetic state of the whole
boson system (i.e. free zone and the localized state) and that leads to the
appearance of a condensate.

The phenomenon of a giant intensity shot was revealed in an
experiment when the IX line shifted to lower energies (that means changing $%
V_{dc} $), this being a consequence of an increase of the radiative lifetime
of excitons due both to an increase of the exciton concentration (at
constant optical pumping) and to their effective thermalization to the bath
temperature. These two circumstances according to the authors opinion cause
the BEC that leaded to a huge population of a localized state.\newline
Since, as noted above, all localized excitons take part in radiative
recombination, this resulted in a significant change of the shape of the IX
luminescence line from the case of figure (c) to that of figure (d).
\begin{figure}
\resizebox{0.48\textwidth}{!}{%
  \includegraphics{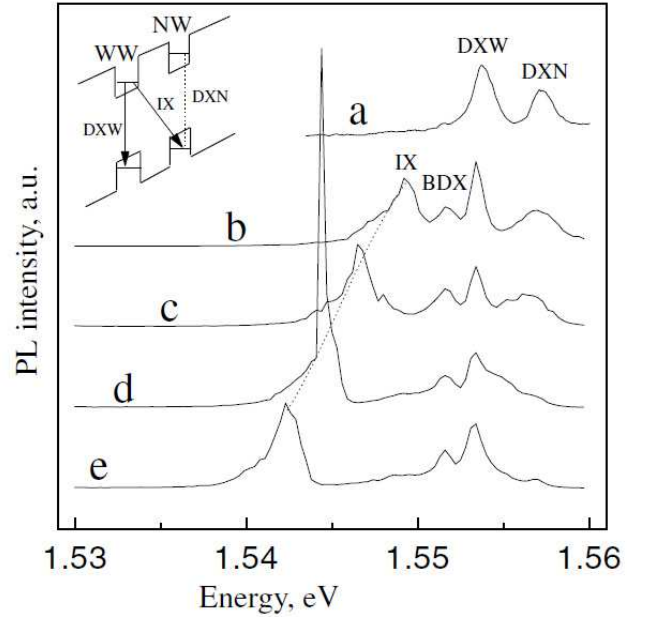}
}
\caption{PL spectra taken at $T=1.8K$, $P=5Wcm^{-2} $ and $V_{dc} =0$ V(a), -0.5 V(b), -2 V(c). The inset
shows the indirect regime of the DQW following the Ref.[142].}
\label{fig:3}       
\end{figure}
\begin{figure}
\resizebox{0.48\textwidth}{!}{%
  \includegraphics{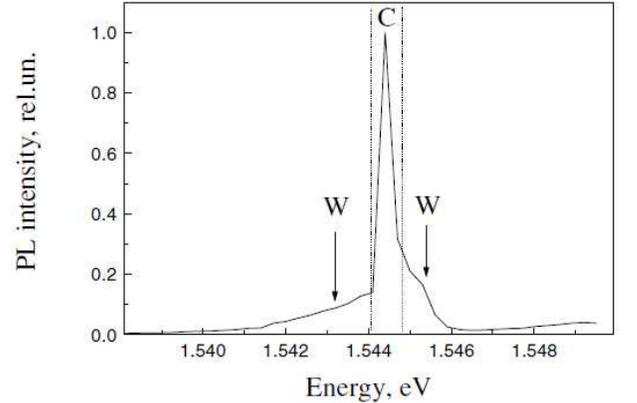}
}
\caption{IX line
spectral profile of figure (d). Two vertical dotted lines separate different
spectral parts of the IX line profile which shown (C) and did not show (W)
temporal evolution of the PL intensity. The data were obtained in Ref.[142].}
\label{fig:4}       
\end{figure}
Thus the totality of the experimental data describing the evolution of the
IX luminescence line (namely, a giant rise of intensity of a part of the
spectral profile of the IX line followed by long-time oscillations) in the
authors opinion provided evidence that in a system of IXs of high density a
Bose--Einstein condensate at localized states (traps) in DQWs appeared.
Thermal equilibrium of this type of excitons in a trap has been demonstrated
experimentally [102, 103].

As was appreciated in the Ref. [102, 103] up to now there has
not been an universally accepted demonstration of BEC in this type of
systems and is necessary a better understanding of the many-body effects of
the interacting dipole IXs. But the accumulated knowledge permits to
formulate some conclusions. One of them states that the confinement of the
excitons in a trap analogous to the optical traps for the cold atoms is a
great advantage instead of creating excitons with a laser allowing them to
expand freely out of the excitation region. Recent work [117] showed that
the IXs in DQWs reach equilibrium both energetically and spatially in a
stress induced trap. One variant of the BEC of excitons equivalent to the
BEC in a trap was proposed by Jan and Lee [149] and was used in Ref. [142,
143]. Another conclusion formulated in Ref. [102. 103] concerns the role of
the temperature. If the temperature is low compared to the energy
fluctuation due to the disorder then the excitons will become trapped in low
energy minima of the disorder potential and will not act as a free gas. Such
energy minima can localize only one or a small number of IXs because they
are repealing each other due to the dipole-dipole interaction. In difference
on it in the trap there are localized levels able to accommodate a
macroscopic number of IXs. A strong energy shift due to interactions may
cancel out the trapping potential and may flatten it [121].

Another important conclusion of Refs. [102, 103] is based on
the results by Laikhtman and Rapaport [151] who underlined that the dipole
IXs in a coupled quantum wells (CQWs) no longer act as a gas but rather as a
correlated liquid. This does not mean that BEC is impossible at high
density. But the canonical telltale for condensation in a weakly interacting
Bose gas, namely a peak of occupation number at $k=0$ may not be easily seen
in these systems. It would be better to look for hydrodynamic effects of
condensation of excitons such as quantum vortices or superfluidity similar
to the liquid helium.

As one can see the most eminent achievements discussed in the
review article concern the famous FQHEs discovered in the frame of the
one-component 2DEG. They suggest to search similar phenomena in the frame of
the two-component, 2D e-h systems, when the CPs will be formed by electrons
and holes with attached point vortices in different combinations taking into
account the interactions inside the CPs from one side and the Coulomb
interaction between the electrons and holes forming the usual
magnetoexcitons from the other side. In spite of the fact that the
supplementary gauge magnetic fields created by electrons and holes with
opposite charges will be compensated in the mean field approximation,
nevertheless the new electron quantum states will appear as elementary
excitations and quantum fluctuations in these new conditions.

\section{Conclusions:}
The purpose of the present review is to discuss the phenomena related to the spontaneous breaking due to the quantum fluctuations of the continuous symmetries existing in the frame of the two-dimensional e-h systems in a strong perpendicular magnetic field with electrons and holes lying on the lowest Landau levels. The spontaneous symmetry breaking leads to the formation of the new ground states and phase transitions and determines the energy spectra of the collective elementary excitations appearing over the new ground states.
	
The main attention is given to the electron-hole systems forming the coplanar magnetoexcitons in the Bose-Einstein condensation ground state with wave vector $\vec{k}=0$ under the influence of the excited Landau levels when the exciton-type excitations coexist with the plasmon-type oscillations. At the same time the properties of the 2DEG under the conditions of the FQHE as well as of the similar 2DHG spatially separated on the layers of the DQW are taken into account, so as to foresee their possible influence on the states of the coplanar magnetoexcitons when the distance between the DQW layers diminishes. Side by side with the 2DEG and 2DHG a bilayer electron systems in the conditions of the FQHE with one half filling factor of LLLs in each layer and with the total filling factor of two layers equal to unity are taken into account because the coherent superposition of the electron states in two layers happens to be equivalent with the formation of the QHExs in the coherent macroscopical state, which can be compared with the BEC of the coplanar magnetoexcitons. The breaking of the global gauge symmetry as well as of the continuous rotational symmetries leads to the formation of the gapless Nambu-Goldstone modes of the collective excitations above the selected ground state, corresponding to the macroscopical wave function with a fixed phase, whereas the breaking of the local gauge symmetry gives rise to the Higgs phenomenon characterized by the gapped branches of the energy spectrum of the collective elementary excitations. The existence of the gapless and gapped branches of the energy spectrum is equivalent to the appearance of the massless and massive particles correspondingly in the relativistic physics.
	
Application of the Nielsen-Chadha theorem establishing the relation between the number of the NG modes and the number of the broken symmetry operators as well as the elucidation of the conditions when the quasi-NG modes appear was effectuated on the concrete example of the spinor atoms in the state of BEC in an optical trap with the aim to better understand the results concerning the coplanar magnetoexcitons. The Higgs phenomenon gives rise to the formation of the composite particles in the frame of the 2DEG in conditions of the FQHE, so that the electron with an odd or even number of the attached point vortices behaves as an composite boson or fermion correspondingly. Their description in the frame of the Ginzburg-Landau theory is demonstrated.
	
Side by side with the 2D coplanar magnetoexcitons the conditions under which the spontaneous coherence may appear in the system of indirect excitations in the double quantum well structures with spatially separated electrons and holes were discussed. The experimental attempts to achieve the BEC of IXs in the traps arising due to the interface width fluctuations or due to the applied stress were reviewed, the concluding remarks and the recommendations are mentioned. The formation of the high density 2D magnetoexcitons and magnetoexciton-polaritons with point quantum vortices attached is suggested.

\end{document}